\documentclass[aps,prl,twocolumn,floatfix,longbibliography,nofootinbib]{revtex4-2}
\usepackage[utf8]{inputenc}
\usepackage{natbib}
\usepackage{graphicx}
\usepackage{xcolor}
\usepackage[tbtags]{amsmath}
\usepackage[colorlinks=true, linkcolor=blue]{hyperref}
\usepackage{amssymb}
\usepackage{gensymb}
\usepackage{comment}
\usepackage{float}
\usepackage{amsmath}
\usepackage{tabularx,graphicx}
\usepackage{epstopdf}
\usepackage{latexsym}
\usepackage{color, colortbl}
\usepackage{psfrag}
\usepackage{bbm}
\usepackage{bm}
\usepackage{titlesec}
\usepackage{dsfont}
\usepackage{feynmp}
\usepackage{slashed}
\usepackage{multirow}
\usepackage{txfonts}
\usepackage{subcaption}
\usepackage{tikz}
\usepackage[newcommands]{ragged2e}
\usepackage[normalem]{ulem}
\renewcommand{\vec}[1]{\boldsymbol{#1}}
\usepackage{braket}
\usepackage{subcaption}

\def \v {{\bf v}}

\def \x {{\vec x}}

\def \beq {\begin{eqnarray}}
\def \eeq {\end{eqnarray}}
\def \tn {\textnormal}

\def \nn {\nonumber}

\newcommand{\rd}{{\rm d}}

\newcommand{\Tr}{{\rm \, Tr\,}}

\newcommand{\calF}{{\mathcal F}}

\renewcommand{\Im}{\text{Im}}

\newcommand{\update}[1]{{ #1}}

\usepackage{pdfpages}

\makeatletter
\AtBeginDocument{\let\LS@rot\@undefined}
\makeatother

\begin{document}
\title{Dynamical Freezing in Exactly Solvable Models of Driven Chaotic Quantum Dots}
\author{Haoyu Guo}
\author{Rohit Mukherjee}
\author{Debanjan Chowdhury}
\affiliation{Department of Physics, Cornell University, Ithaca, New York 14853, USA.}
\begin{abstract}
The late-time equilibrium behavior of generic interacting models is determined by the coupled hydrodynamic equations associated with the globally conserved quantities. In the presence of an external time-dependent drive, non-integrable systems typically thermalize to an effectively infinite-temperature state, losing all memory of their initial states. However, in the presence of a large time-periodic Floquet drive, there exist special points in phase-space where the strongly interacting system develops approximate {\it emergent} conservation laws. Here we present results for an exactly solvable model of two coupled chaotic quantum dots with multiple orbitals interacting via random two and four-fermion interactions in the presence of a Floquet drive. We analyze the phenomenology of dynamically generated freezing using a combination of exact diagonalization, and field-theoretic analysis in the limit of a large number of electronic orbitals. The model displays universal freezing behavior irrespective of whether the theory is averaged over the disorder configurations or not.
We present explicit computations for the growth of many-body chaos and entanglement entropy, which demonstrates the long-lived coherence associated with the interacting degrees of freedom even at late-times at the dynamically frozen points. We also compute the slow timescale that controls relaxation away from exact freezing in a high-frequency expansion.
\end{abstract}

\maketitle

{\it Introduction.-} The quantum statistical mechanics of complex many-body Hamiltonians presents a fertile ground for exploring novel phenomena in both theory and experiment. Key questions revolve around thermalization dynamics \cite{Deutsch,srednicki,Tasaki,rigol,Greiner,schmied,Blatt14,APreview} in generic quantum systems undergoing unitary evolution. While non-integrable, translationally invariant Hamiltonians are expected to thermalize, strong disorder can possibly lead to many-body localization (MBL) \cite{basko2006metal,gornyi,Huserev,Altmanrev,RMPMBL}. The stability of MBL as a phase of matter in the thermodynamic limit remains debated \cite{roeck,PhysRevB.99.134305,vidmar,avalanches,sels}. External drives, like time-periodic Floquet drives, effectively thermalize to infinite temperature in non-integrable models \cite{PhysRevE.90.012110,PhysRevX.4.041048}, erasing all initial state information. Recent research highlights Floquet MBL and time-crystalline order as mechanisms to resist heating in strongly disordered systems \cite{Sacha_2018,DTC_AR,DTCRMP}. Quantum ``scar" states offer another avenue for retaining memory of initial states, yet they are sparse within the many-body spectrum of generic Hamiltonians \cite{Serbyn2021,Scars_AR}.

Various translationally invariant quantum spin-models in one and two dimensions have been proposed to exhibit ``freezing" under large Floquet drives, characterized by specific combinations of drive amplitude and frequency \cite{PhysRevB.82.172402,PhysRevB.86.054410,PhysRevB.90.174407,PhysRevB.97.245122,PhysRevX.11.021008,KS1,KS2,KS3,KS4,KS5,KS7,sreemayee}. This dynamical freezing entails {\it approximate emergent} conservation laws associated with macroscopic observables {\it not} conserved in the static Hamiltonian. The emergent conserved observable depends on the drive Hamiltonian. For instance, in a driven Ising-type quantum spin model without any conserved quantities, the Floquet drive $f(t)\sum_i\sigma_x$ can induce a conserved magnetization $(\sum_i\sigma_x)$ at specific phase-space points. Investigations of freezing have relied on exact diagonalization, supplemented by a {leading-few-order} Magnus-type expansion (which we review below). {Previous analyses have not determined if freezing is an approximate phenomenon, nor identified what controls its time scale if it is.} It would be useful to have complementary insights into dynamical freezing, extend the phenomenology to other settings , and diagnose its other salient features.

  {Here we examine a ``solvable" model involving two coupled quantum dots with $N$ orbitals each, described by the Sachdev-Ye-Kitaev model \cite{SSachdev1993,AYKitaev2015,JMaldacena2016c,DChowdhury2022a,CKuhlenkamp2020,PhysRevResearch.6.013094,PZhang2019,ALarzul2022,YCheipesh2021}, which becomes analytically tractable in the limit of $N\to\infty$ orbitals  {through field-theoretic techniques}. Notably, our setup differs from prior studies, expanding the realm of the freezing phenomena beyond spin chain systems. {Most interestingly, while previously studied examples host well-defined quasiparticles in the static limit, the example considered here hosts no quasiparticle-like excitations and have short equilibration times. Whether such many-body systems can be frozen even approximately is {\it a priori} unclear.} We demonstrate that {\it strongly correlated} systems with a local Hilbert space dimension approaching $N\rightarrow\infty$ and {\it without} the notion of spatial locality exhibit freezing. Furthermore, freezing occurs not only at the level of individual disorder realizations studied via exact diagonalization methods, but also persists after disorder averaging. As a new characterization of freezing, we find that many-body chaos can be mitigated dynamically via Floquet drive.  Our setup also offers a unique technical advantage, which can be investigated both through exact diagonalization for small-$N$, and Schwinger-Dyson equations in the large-$N$ limit.} {In particular, the large-$N$ field-theoretic formulation has allowed us to quantitatively compute the corrections to the freezing phenomenology, which goes beyond the Magnus expansion and reveals the approximate nature of the freezing phenomenology.} In a related work \cite{transmon}, we have also studied a driven model of coupled transmons — a prominent qubit hardware for quantum computing applications \cite{cqed} — which also exhibits dynamical freezing.

\begin{figure*}[htb]
  \centering
  \begin{subfigure}[t]{0.18\textwidth}
      \includegraphics[width=\textwidth]{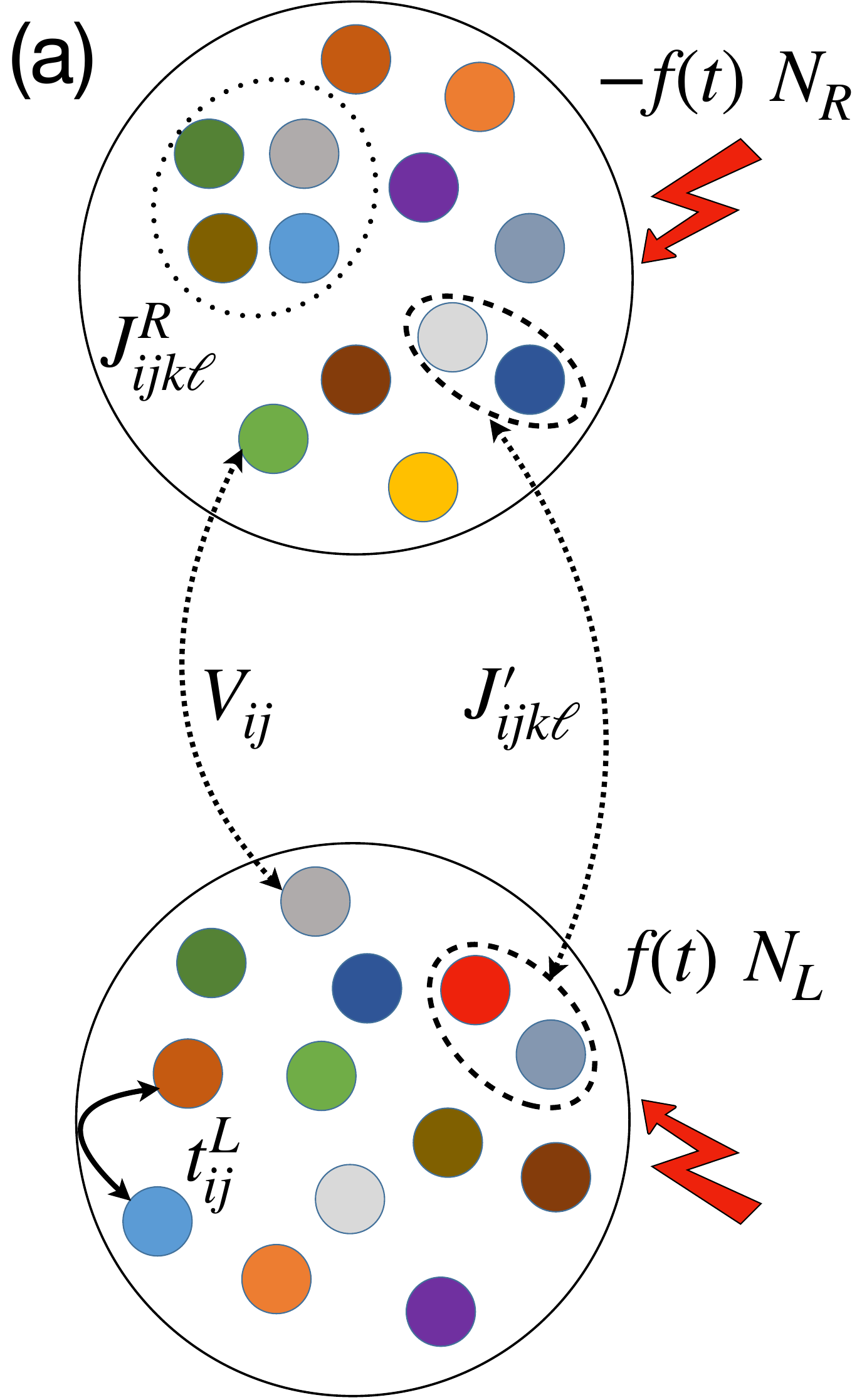}
  \end{subfigure}
  \begin{subfigure}[t]{0.35\textwidth}
      \includegraphics[width=1.0\linewidth]{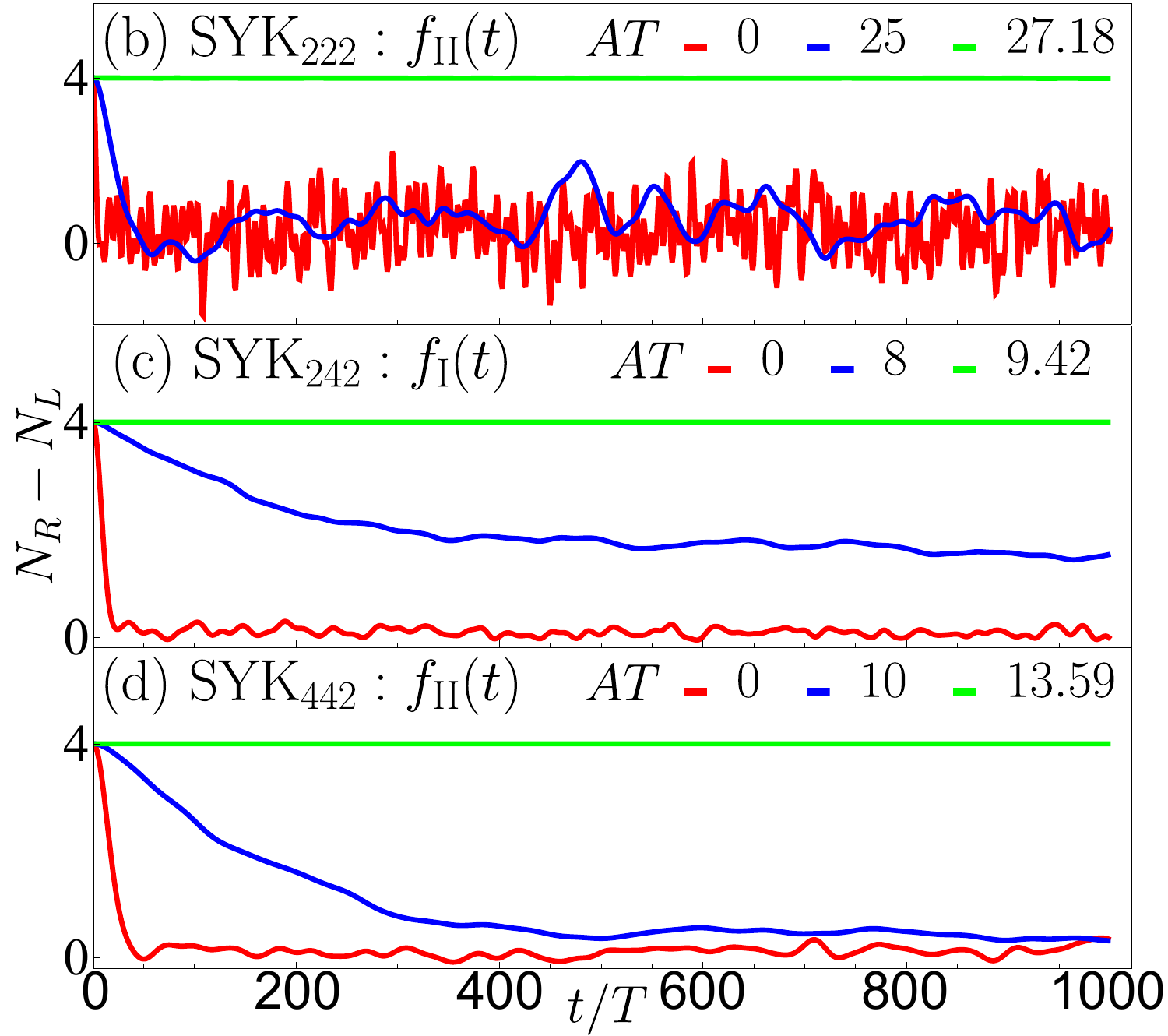}
  \end{subfigure}
  \begin{subfigure}[t]{0.35\textwidth}
      \includegraphics[width=1.0\linewidth]{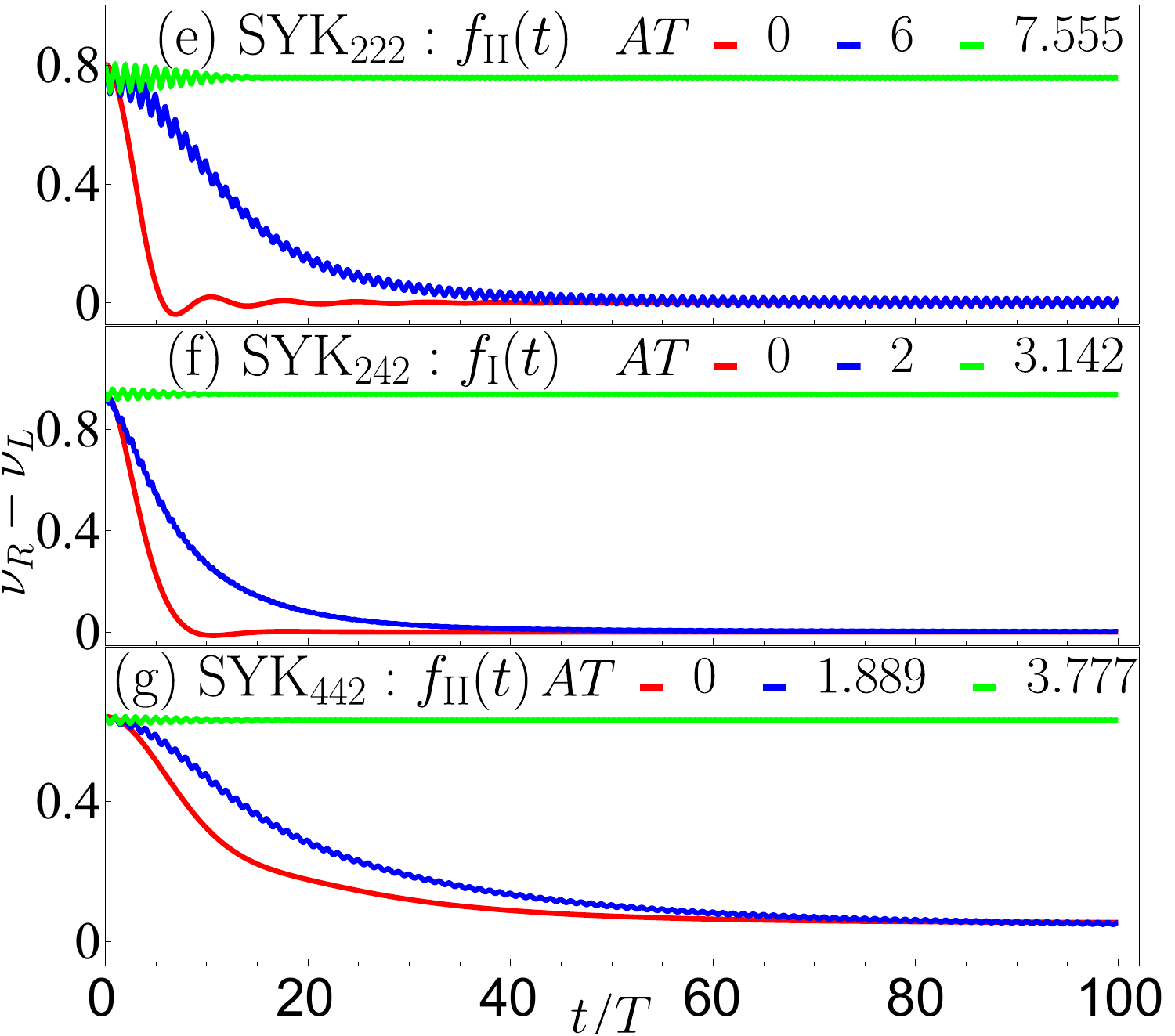}
  \end{subfigure}
  \caption{(a) Illustration of two coupled SYK quantum dots with static coupling $t,J,V,J'$; the density difference $N_L-N_R$ is driven via a function, $f(t)$ [Eq.~\ref{eq:drives}]. (b)-(d) ED results for the difference in particle numbers of two dots, $N_R(t)-N_L(t)$, plotted stroboscopically. For each system, we show three different driving amplitudes: undriven (red), driven away from freezing (blue) and driven at the freezing point (green). The system parameters are: (b) $\rm{SYK}_{222}$ with $t_LT=t_RT=(1/2)V_{LR} T=0.1$ with the drive in Eq. \eqref{drive2}; (c) $\rm{SYK}_{242}$ with $t_LT=t_RT=(1/2)J_{LR}T=0.1$ with the drive in Eq.~\eqref{drive1}; (d)  $\rm{SYK}_{442}$ with $J_L T=t_R T=J_{LR}T $ $=0.1$ with the drive in Eq.~\eqref{drive2}. The ED results are obtained with $N=6$ and initial condition $N_R=5$, $N_L=1$, for individual disorder realizations.
  (e)-(g) Numerical results of KBE for the difference in density of two dots $\nu_R(t)-\nu_L(t)$.
  The system parameters are the same as in ED. We initialize the system in a low-temperature thermal state with inverse temperature $\beta=50 T$ ($T\equiv$period of the Floquet drive), and chemical potential $\mu_R T=-\mu_L T=0.25$ (e,f) or $0.06$ (g).
  }\label{cartoon}
\end{figure*}

{\it Model.-} We will be interested in a model of two coupled $(0+1)-$dimensional interacting quantum dots  (Fig.~\ref{cartoon}a), labeled as $\alpha = L,~R$, with a Hamiltonian $H = H_{\rm{static}} + H_{\rm{dynamic}}$:
\begin{subequations}
\beq
H_{\rm{static}} &=& H_L + H_R + H_{LR},\\
H_{\alpha = \{L,R\}} &=& \sum_{i,j} \left(t^\alpha_{i,j}-  {\mu_\alpha\theta(-t)}\delta_{ij}\right)c_{\alpha,i}^\dagger c^{\phantom\dagger}_{\alpha,j} + \sum_{i,j,k,\ell} J^\alpha_{ijk\ell} c^\dagger_{\alpha i}c^\dagger_{\alpha j} c^{\phantom\dagger}_{\alpha k} c^{\phantom\dagger}_{\alpha\ell} , \nn \\
H_{LR} &=& \sum_{i,j} V_{ij} c_{L,i}^\dagger c^{\phantom\dagger}_{R,j} +  \sum_{i,j,k,\ell} J'_{ijk\ell} c^\dagger_{L i}c^\dagger_{L j} c^{\phantom\dagger}_{R k} c^{\phantom\dagger}_{R\ell} + \tn{h.c.}, \nn \\
H_{\rm{dynamic}} &=&  {\theta(t)}f(t) \sum_i\bigg[c^\dagger_{L,i}c^{\phantom\dagger}_{L,i} - c^\dagger_{R,i}c^{\phantom\dagger}_{R,i}\bigg] \equiv  {\theta(t)}f(t) H_{\rm{drive}}, \label{eq:drive}
\eeq
\end{subequations}
 {where $\theta(t)$ is the standard Heaviside function.} Here, we have assumed both dots have  $i,j=1,..,N$ orbitals. The couplings, $t_{i,j}^{\alpha}=\left(t_{j,i}^{\alpha}\right)^*$ and $J_{ijk\ell}^{\alpha}=-J_{jik\ell}^{\alpha}=-J_{ij\ell k}^{\alpha}=\left(J^\alpha_{k\ell ij}\right)^*$ are the intra-dot tunneling and interaction matrix elements, respectively, drawn from independent Gaussian distributions with zero mean,
and finite variances, $\overline{|t_{i,j}^{\alpha}|^2}=(t_\alpha)^2/N,~\overline{|J^\alpha_{ijk\ell}|^2}=(J_{\alpha})^2/(2N)^3$. Note that the two dots need not be identical.
 The two dots are coupled via a single-particle tunneling, $V_{ij}$, and a two-particle tunneling, $J'_{ijk\ell}=-J'_{jik\ell}=-J'_{ij\ell k}$, that are also drawn from independent Gaussian  distributions with zero mean,
and finite variances, $\overline{|V_{i,j}|^{2}}=V_{LR}^2/N,~\overline{|J'_{ijk\ell}|^2} = J_{LR}^2/(2N)^3$. In the absence of a coupling between the two dots, $H_{\rm{static}}$ exhibits an {\it exact} $U(1)\times U(1)$ symmetry tied to the independently conserved densities, $N_{\alpha=L,R}=\sum_i c_{\alpha,i}^\dagger c^{\phantom\dagger}_{\alpha,i}$, which can be tuned by the chemical potential, $\mu_\alpha$. However, even for an infinitesimal $H_{LR}$, this reduces to a single $U(1)$ symmetry tied to the total density, $N_\text{tot} = N_L + N_R$.
   We will analyze three special cases, denoted SYK$_{pqr}$:
(i) SYK$_{222}$, a fully integrable random-matrix model with intra-dot SYK$_2$ and inter-dot SYK$_2$ hoppings (i.e. $J_L,J_R,J_{LR}=0$), (ii) SYK$_{242}$, an intra-dot SYK$_2$ model coupled via SYK$_4$-type inter-dot interactions (i.e. $J_L,J_R,V_{LR}=0$), and (iii) SYK$_{442}$, an SYK$_4$ dot coupled to an SYK$_2$ dot via SYK$_4$-type interactions (i.e. $t_L,J_R,V_{LR}=0$).
 {When $t<0$, we prepare the system with a chemical potential bias $\mu_R\neq \mu_L$. At $t=0$, we remove the bias and switch on the Floquet drive for all $t>0$.}

The remainder of this article is concerned with the properties of the above Hamiltonian with a periodic drive, $f(t)=f(t+T)$, where $T$ represents the  period (we denote temperature by $\beta^{-1}$). We concentrate on two specific drives:
\begin{subequations}\label{eq:drives}
\beq
{\rm{Case~ I:}}~~f_{\rm{I}}(t) &=& A~{\rm{sgn}}[\sin(\Omega t)],\label{drive1}\\
{\rm{Case~ II:}}~~f_{\rm{II}}(t) &=& A\cos(\Omega t),\label{drive2}
\eeq
\end{subequations}
where $A$ denotes the drive amplitude and $\Omega=2\pi/T$ is the frequency.
Initially, we conduct an exact diagonalization analysis for small $N$, examining both disorder-averaged and non-averaged results; the freezing behavior is qualitatively independent of averaging \cite{supp}. Subsequently, we delve into the disorder-averaged non-equilibrium Kadanoff-Baym equation in the large-$N$ limit, which we solve numerically. Both approaches yield nearly identical outcomes, providing complementary insights into freezing physics. Importantly, neither $A$ nor $\Omega$ needs to surpass the $O(N)$ many-body bandwidth.
 As we will illustrate, the freezing point is characterized by an approximate emergent conservation law, where $N_R$ and $N_L$ are conserved separately. The drive effectively ``decouples" the $L$ and $R$ dots dynamically. Furthermore, once $H_{LR}$ is effectively decoupled, the structure of many-body chaos at freezing for the SYK$_{pqr}$ models is governed by the chaotic properties of the individual dots described by $H_{L}$ and $H_R$, respectively.

{
{\it General criterion for dynamical freezing.-} We first present a necessary, but not sufficient condition for dynamical freezing based on a ``generalized" Magnus expansion \cite{AHaldar2020}, which treats the $\Omega\to\infty$ limit while keeping $A/\Omega$ finite. The procedure first goes to a rotating reference frame before performing the conventional Magnus expansion, which yields the following condition for freezing at leading order (see the End Matter for detailed derivation):
\begin{subequations}\label{eq:criterion}
\beq\label{}
  AT(E_m-E_n) &=& 4\pi\mathbb{Z}\,,\quad\forall ~m,n,~~ ({\rm Case~I}),\\
   AT\left(E_m-E_n\right) &=& 2\pi\zeta_i,\quad\forall ~m,n,~~ ({\rm Case~II}),
\eeq
\end{subequations}
where $\zeta_i$ are zeroes of the Bessel function $J_0(\cdot)$.  

We note that the condition above does not guarantee freezing, especially when the many-body spectrum of the static Hamiltonian is as chaotic and unconventional as in the SYK models. However, if the freezing phenomenon does occur, we expect its position will be characterized by Eq.\eqref{eq:criterion}, especially in the limit of $\Omega\gg \{t_\alpha,J_\alpha,V_{LR},J_{LR}\}$, i.e. the energy scale of single-particle or two-particle excitations. Crucially, the discussion so far has not relied on either the large-$N$ limit or disorder-averaging. Therefore, if dynamical freezing arises in the regime suggested above, it should do so at small$-N$ for individual disorder realizations, as well as in the large$-N$ disorder-averaged limit. We also point out that freezing is rendered approximate by terms beyond the leading-order Magnus analysis here, which we will demonstrate later by the nonzero conductivity between the two quantum dots.}

 {\it Emergent conservation: Exact diagonalization.-} To test the possibility of an emergent (approximate) conservation associated with the density difference $N_R-N_L$, we apply the criterion in Eq.~\eqref{eq:criterion} to the coupled SYK dots. \update{We illustrate the appearance of dynamical freezing for only three different combinations of $\rm{SYK}_{pqr}$ and a Floquet drive; however, any combination of values of $p,~q,~r$ and drive-protocols realize freezing determined by Eq.~\eqref{eq:criterion}. Specifically,} we study (i) $\rm{SYK}_{222}$ with cosine drive $f_{\rm II}$; (ii) $\rm{SYK_{242}}$ with square-wave drive $f_{\rm I}$; and (iii) $\rm{SYK_{442}}$ with cosine drive $f_{\rm II}$. For an ${\rm{SYK}}_q$ inter-dot interaction ($q=2,~4$), $E_m-E_n=\pm q$ for any $m,n$ such that $\braket{m|H_\text{static}|n}\neq 0$. Hence the expected freezing points for the above cases are (i) $AT=\pi \zeta_i$; (ii) $AT=\pi k$ for $k\in\mathbb{Z}_+$; and (iii) $AT=\pi \zeta_i/2$. \update{We note that there is no particular reason behind choosing these specific combinations of interactions and drive-protocols, and the qualitative features remain similar.}

We begin by analyzing the physics at small-$N$, where we use exact diagonalization employing QuSpin \cite{PWeinberg2019} to calculate the stroboscopic time-evolution of an initial  {pure state} $\ket{\psi}$ for a \emph{single} disorder realization of $H$.  {
We initialize the system in a thermal superposition of the eigenstates of the operator $\hat{N}_{D}=\hat{N}_{R}-\hat{N}_{L}$, $\ket{\psi}=\frac{1}{\mathcal{N}}\sum_{n}e^{-\beta E_{n}}\ket{n}$, where $E_{n}$ and $\ket{n}$ are the eigenvalues and eigenvectors of $\hat{N}_{D}$ respectively, $\mathcal{N}$ is the normalization constant.
} As shown in Fig.~\ref{cartoon}(b)-(d), for the undriven system ($AT=0$) as well as when the driven system is away from any freezing point, the density difference $(N_R-N_L)$ rapidly decays away from its initial value towards zero. The small fluctuations around this asymptotic equilibrated value is expected \cite{APreview}, and is especially pronounced in the quadratic $\rm{SYK}_{222}$ system due to additional coherent quantum interference effects. On the other hand, when the system is tuned to the freezing points based on Eq.~\eqref{eq:criterion}, the decay of $N_R-N_L$ is strongly suppressed and remains approximately conserved up to late times. Notably, every single disorder realization of the Hamiltonian and arbitrary initial states, including thermal states at high temperatures, displays freezing \cite{supp}. To gain further analytical insights into the structure of freezing, we turn next to the large-$N$ limit.

{\it Emergent conservation: Large$-N$ analysis.-} We will analyze the large-$N$ limit using  Kadanoff-Baym equations (KBE) \cite{KBoriginal,GStefanucci2013a}. At the leading order, the system is described by Green's functions,  $G_{\alpha}(z,z')=-i(1/N)\sum_i T_C\braket{c_{\alpha i}(z)c_{\alpha i}^\dagger(z')}$, where $z,z'$ are coordinates on the ``L-shaped" Keldysh contour $C$ (consisting of both real and imaginary-time branches) \cite{GStefanucci2013a,MSchuler2020}, and $T_C$ denotes path ordering on $C$. The KBE is given by \cite{MSchuler2020,GStefanucci2013a}
\begin{subequations}\label{eq:KBE}
    \begin{eqnarray}
    i\partial_z G_\alpha(z,z')&=&\int_C\rd \bar{z}~\Sigma_\alpha(z,\bar{z})~G_\alpha(\bar{z},z')\,, \\
 -i\partial_{z'} G_\alpha(z,z')&=& \int_C \rd\bar{z} ~G_\alpha(z,\bar{z})~\Sigma_\alpha(\bar{z},z')\,.
    \end{eqnarray}
\end{subequations}
Note that we have encoded the driving term into the self-energy $\Sigma$, after a gauge-transformation. The self-energy reads
\begin{align}\label{eq:self_energy}
  &\Sigma_\alpha(z,z')=t_\alpha^2 G_\alpha(z,z')+J_\alpha^2 G_\alpha^2(z,z')G_\alpha(z',z)\\
  &+V_{LR}^2 e^{2i\Theta(z)}G_{\bar{\alpha}}(z,z')e^{-2i\Theta(z')}+J_{LR}^2 e^{4i\Theta(z)}G_{\bar{\alpha}}^2(z,z')G_\alpha(z',z) e^{-4i\Theta(z')}.\nonumber
\end{align}
Here $\Theta(z)=\int_0^z \rd t f(t)$ is the integral of the drive function (see also Eq.\eqref{eq:theta}). We solve Eqs.~\eqref{eq:KBE} and \eqref{eq:self_energy} numerically using the NESSi package \cite{MSchuler2020,supp}. The system is  initialized in a thermal state with unequal chemical potentials on the two dots, and we switch on the drive at $t=0$.
 The densities can be extracted through the lesser Green's function $\nu_\alpha(t)=-iG_\alpha^{<}(t,t)=(1/N)\sum_i\braket{c_{\alpha i}^\dagger(t) c_{\alpha i}(t)}$.
As shown in Fig.~\ref{cartoon}(e)-(g), the density difference $\nu_R-\nu_L$ at freezing remains approximately conserved up to late times, and rapidly equilibrates
to zero away from freezing.  {By analyzing the single-particle spectrum in the presence of the drive, we find that at ``low-(quasi)energy" the two dots satisfy fluctuation-dissipation relation with a local effective temperature and chemical potential, respectively \cite{supp}. At freezing, once the two dots are dynamically decoupled, they fail to equilibrate with each other upto late times.}

We note that while the freezing dynamics is similar between the large-$N$ KBE and small-$N$ ED, the non-freezing dynamics show some difference. The non-freezing ED results show stronger fluctuations, likely related to the single-disorder snapshots and finite-size effects \cite{PhysRevE.89.042112,PhysRevLett.123.010601}.

\begin{figure}[tb]
  \centering
  \includegraphics[width=0.9\linewidth]{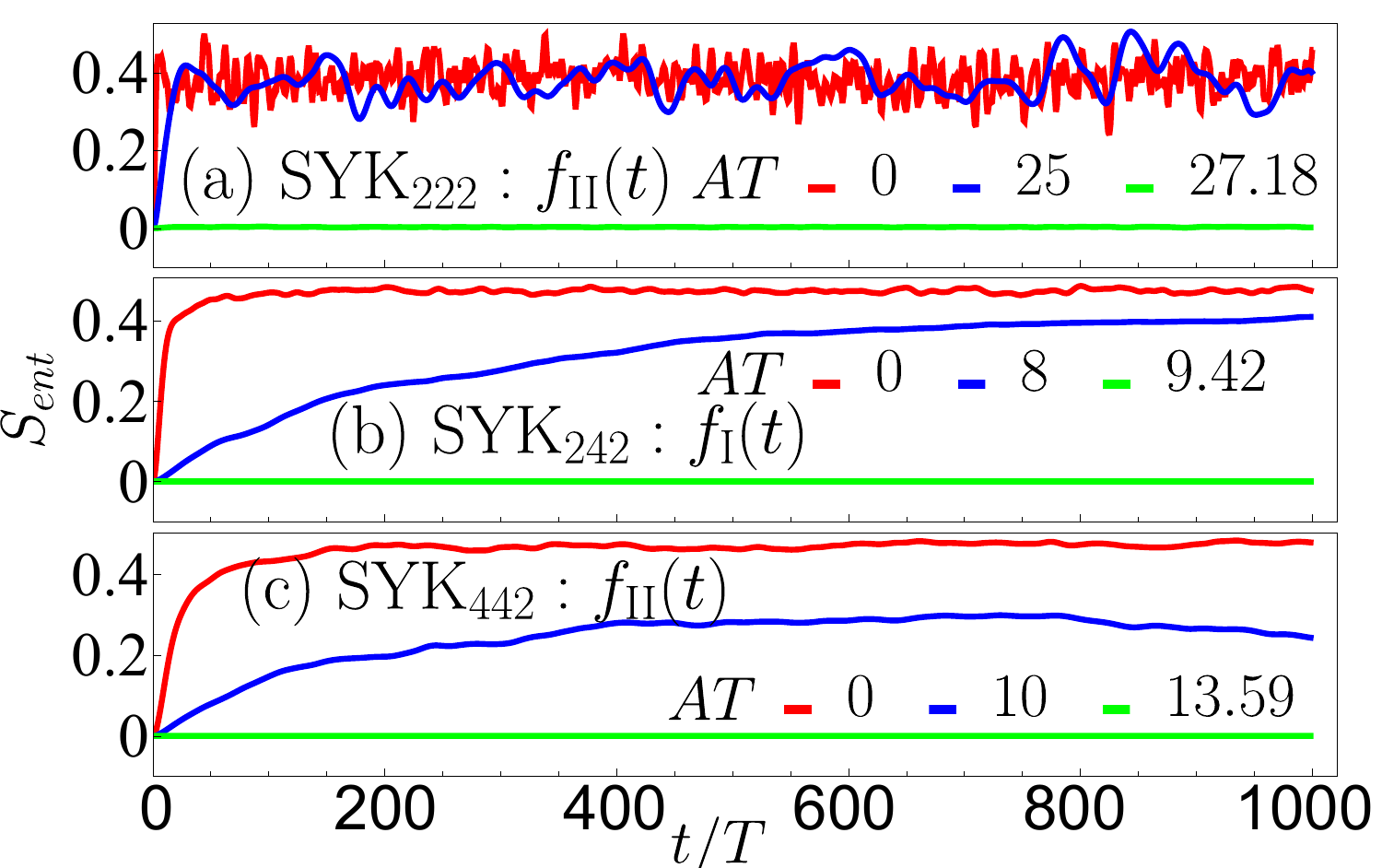}
  \caption{Stroboscopic evolution of entanglement entropy ($S_{ent}$) at and away from freezing. Here $T$ is the period of the Floquet drive.
  Plots are generated for individual snapshots of disorder  {and the parameters are same as in Fig.~(\ref{cartoon}) with the same initial state.}}\label{fig:EDSent}
\end{figure}

{\it Dynamical decoupling, Entanglement growth, and Chaos.-}
To help distinguish the nature of the residual quantum dynamics at and away from freezing, we focus on two different diagnostics. First, we compute the entanglement entropy (EE) stroboscopically of a single dot $S_\text{ent}=-\Tr_R\rho_R\ln\rho_R$, $\rho_R=\Tr_L \rho_{LR}$ using exact-diagonalization for the small systems, as before. Next, we analyze the growth of chaos in the large$-N$ framework by extending the KBE-based formalism to account for the additional backward time-evolution. {We note that a generic characterization of quantum chaos in floquet systems is still an open question \cite{CLim2024,SBhattacharjee2024,TLeBlond2021}, and here the large-$N$ formulation has allowed us to compute the out-of-time-order correlators (OTOC) which is an indicator of chaos.}

The results for EE are shown in Fig.~\ref{fig:EDSent}. We initialize the system in a product state, such that the two dots are unentangled. Away from freezing, we find that the EE grows and saturates. On the other hand, at freezing the EE does not display any appreciable growth up to late times.

Turning to the large-$N$ system, we put the system on a double Keldysh contour $C_d$ with two copies of (real-time) Keldysh contours $C_d=C_1\cup C_2$ \cite{supp}. The diagnosis for chaos is the inter-branch Green's function $G^{12}(z_1,z_2)$ where $z_1\in C_1$ and $z_2\in C_2$. In a chaotic many-body system, especially for models with a large$-N$ or weakly-coupled semiclassical limit, $G^{12}(t,t)$ grows expnentially $G^{12}(t,t)-G^<(t,t)\propto\exp(\lambda_L t)$ where $\lambda_L$ is the Lyapunov exponent of quantum chaos \cite{YGu2022, ILAleiner2016a,YGu2022,PZhang2021a,AKamenev2023,Patel:2016wdy,Patel:2017vfp,Stanford:2015owe,Chowdhury:2017jzb,Steinberg:2019uqb,Grozdanov:2018atb,Kim:2020jpz,Keselman:2020fmo}.  $G^{12}(z_1,z_2)$ is the solution of the modified KBE on $C_d$ with the following perturbation to the self-energy \cite{YGu2022}:
\begin{equation}\label{eq:se_pert}
  \delta\Sigma_{\alpha}(t_1,t_2)=\delta(t_1)\delta(t_2)\times \left\{
\begin{array}{ll}
   -iz, & \hbox{if }t_1\in C_1,t_2\in C_2\,, \\
  iz, & \hbox{if } t_2\in C_1,t_1\in C_2\,.
     \end{array}
 \right.
\end{equation} Here $z$ is a small perturbation that trigers the exponential growth. We note that $G^{12}$ can be related to OTOC via a gluing formula \cite{YGu2022}, and they share the same early-time growth with exponent $\lambda_L$.

\begin{figure}
  \centering
  \includegraphics[width=0.9\columnwidth]{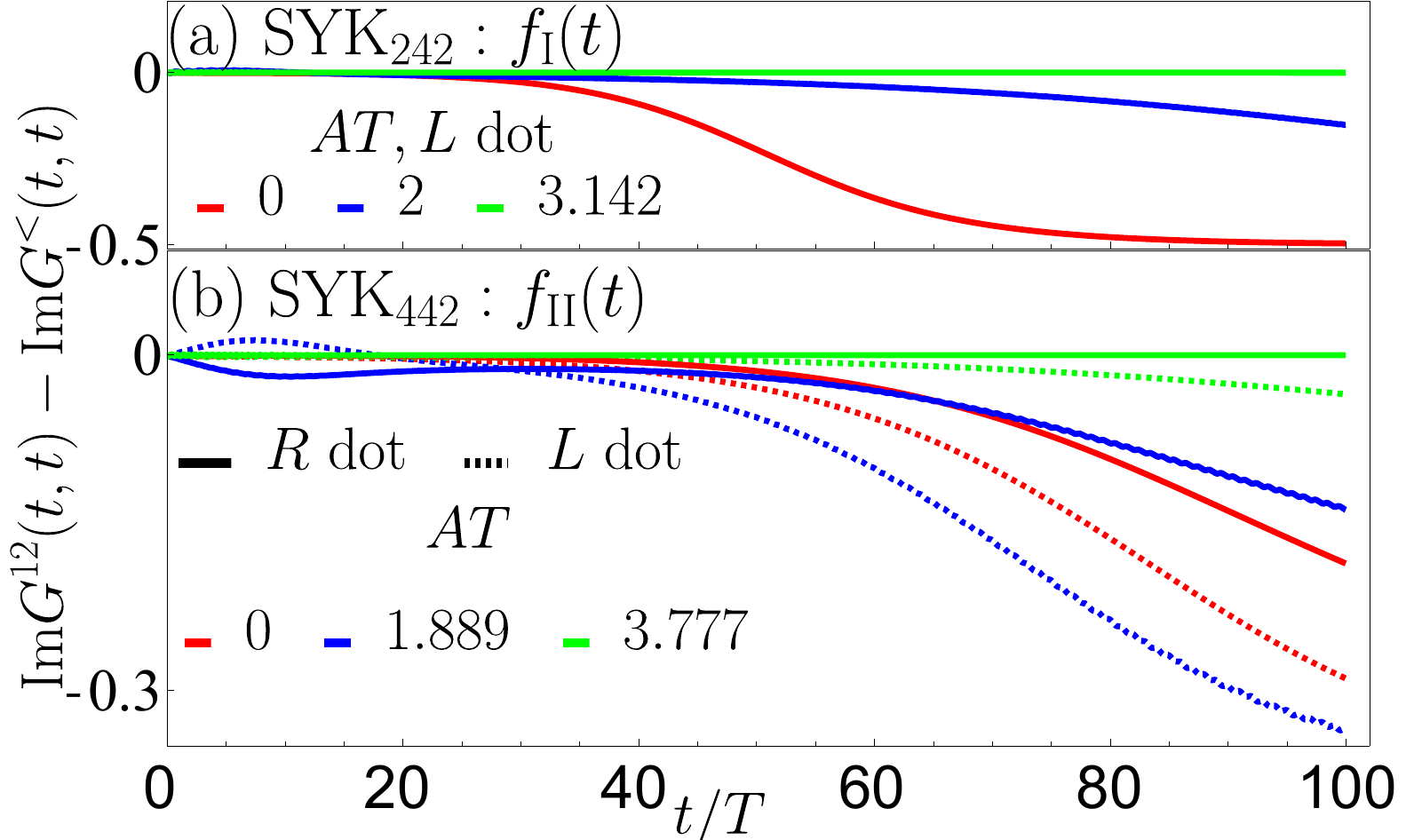}
  \caption{Quantum chaos diagnosed by $G^{12}(t,t)-G^<(t,t)$. At $t=0$, $G^{12}$ is perturbed from $G^<$ by Eq.~\eqref{eq:se_pert} with $z=0.001$. Here $T$ is the period of the Floquet drive. (a) $\rm{SYK}_{242}$ system as in Fig.~\ref{cartoon}(f). Only data for $L$ dot is shown and the $R$ dot behaves similarly. (b) $\rm{SYK}_{442}$ system as Fig.~\ref{cartoon}(g). At the freezing point, the left $\rm{SYK}_4$ dot remains chaotic, but right $\rm{SYK}_2$ dot is integrable. }\label{fig:otoc}
\end{figure}

 We solve the modified KBE in Eq.~\eqref{eq:KBE} with the self-energies in Eq.~\eqref{eq:self_energy} and \eqref{eq:se_pert} numerically \cite{supp}. We present results for quantum chaos in Fig.~\ref{fig:otoc} for $\rm{SYK}_{242}$ and $\rm{SYK}_{442}$; we skip $\rm{SYK}_{222}$ given its non-interacting integrable nature.  When the system is away from freezing and for both models, the two dots are coupled together by the chaotic $\rm{SYK}_4$ interaction, as  reflected in the growth of $G^{12}(t,t)-G^<(t,t)$.
 In  $\rm{SYK}_{242}$, the non-integrability only arises due to the inter-dot $J_{LR}$ coupling. At freezing, both dots effectively decouple into two SYK$_2$, which do not harbor any many-body chaotic behavior, as shown in Fig.\ref{fig:otoc}(a). On the other hand, for $\rm{SYK}_{442}$ at the freezing point upon decoupling, the $R$ dot is an isolated $\rm{SYK}_2$ while the $L$ dot is an isolated SYK$_4$. Therefore, the chaos for the $R$ dot should reflect the integrability while the chaos for the $L$ dot should remain chaotic. This is indeed observed in Fig.\ref{fig:otoc}(b). Therefore, at freezing once the inter-dot coupling $J_{LR}$ is effectively decoupled, the remaining chaotic properties are determined by the intra-dot interactions.

{\it Leading order correction to freezing.-}
The emergent conservation law at the freezing point is only approximate. This immediately leads to the important question of what controls the timescale of the slow non-conserving processes, which our large$-N$ analysis can address analytically.
We analyze the Schwinger-Dyson (SD) equations in the Floquet steady state within a large $\Omega$ expansion. The Green's functions possess discrete time translation symmetry $G(t,t')=G(t+T,t'+T)$, with associated fourier transform
\begin{equation}\label{}
  \hat{G}_{mn}(\omega)=\int_0^T \frac{\rd t }{T} \int_{-\infty}^{\infty} \rd t'G(t,t')e^{-i(\omega+m\Omega)t+i(\omega+n\Omega)t'}\,.
\end{equation} There is a redundancy $\hat{G}_{mn}(\omega+\Omega)=\hat{G}_{m+1,n+1}(\omega)$, which we remove by restricting $\omega\in (-\Omega/2,\Omega/2]$. Here $G$ denotes the retarded or the lesser Green's function.
The SD equation reads
\begin{subequations}\label{eq:SDOmega}
    \begin{eqnarray}
    \hat{G}_\alpha(\omega)&=&\left(\omega+\hat{m}\Omega-\hat{\Sigma}_\alpha(\omega)\right)^{-1}\,,\label{eq:SDF1}\\
    \hat{\Sigma}_\alpha(\omega)&=& \hat{\Sigma}_{\alpha}^{\text{intra}}(\omega)+ \hat{F}_\alpha \hat{\Sigma}_{\alpha}^{\text{inter}}(\omega) \hat{F}_\alpha^\dagger\,, \label{eq:SDF2}
\end{eqnarray}
\end{subequations}
 where $\alpha=L,R$ and $\hat{m}_{ab}=a\delta_{ab}={\rm diag}(\dots,-1,0,1,\dots)$, with $\hat{F}_L=\hat{F}_R^\dagger=\hat{F}$ a unitary matrix with $\hat{F}_{mn}=F_{m-n}$, and $F_m$ is the fourier coefficient of $e^{iq\Theta(t)}=\sum_{m} F_m e^{-im\Omega t}$. Here $\Theta(t)$ is as defined in Eq.\eqref{eq:theta}, and $q$ is the order of inter-dot SYK interaction. $\hat{\Sigma}^{\text{intra/inter}}$ describes intra- and inter- dot interactions (i.e. the first and the second line of Eq.\eqref{eq:self_energy}), respectively.

Since we are interested in the dynamics of the system at times much longer than the period $T$, we can set $\omega\ll \Omega$. Next, we assume the driving frequency $\Omega$ to be higher than the single-particle energy scales, so $\Im \hat{\Sigma}(\pm \Omega/2)\approx 0$. Therefore, at the leading $O(1/\Omega)$ order, $\hat{G}_{00}$ and $\hat{\Sigma}_{00}$ are approximately decoupled from the rest of the matrix, which is much smaller in $1/\Omega$, so the SD equation is approximately,
\begin{subequations}\label{eq:SD0}
\begin{eqnarray}
    \hat{G}_{\alpha,00}(\omega) &=& \left(\omega-\hat{\Sigma}_{\alpha,00}(\omega)\right)^{-1}\,, \\
   \hat{\Sigma}_{\alpha,00}(\omega) &=& \hat{\Sigma}_{\alpha,00}^{\text{intra}}(\omega)+ |F_0|^2\hat{\Sigma}_{\alpha,00}^{\text{inter}}(\omega) \,.
\end{eqnarray}
\end{subequations}
Note that the freezing condition Eq.~\eqref{eq:criterion} exactly corresponds to $F_0=0$, where the two dots become approximately decoupled.

The degree of non-conservation associated with the individual $N_L,~N_R$ can be quantified by the inter-dot conductance $\sigma$ between the two dots averaged over one drive period.
To estimate $\sigma$, we analyze the higher order terms in $1/\Omega$ in Eq.~\eqref{eq:SDOmega}.
At freezing, the effective inter-dot coupling does not appear in the first two orders in $1/\Omega$.
We find that for the $\rm{SYK}_{222}$ system, the leading correction to the conductance is
\begin{equation}\label{eq:sigma222}
  \sigma = \frac{V_{LR}^2 t_L t_R}{\Omega^4} \calF_{222}(V_{LR}/t_{L},t_R/t_L,T_\text{eff}/t_L)\,.
\end{equation} Here $\calF_{222}$ is a scaling function that is not fixed by our calculation. $T_\text{eff}$ is an effective temperature of the system when it freezes \cite{supp}, which is in general different from the initial temperature due to the transient heating in reaching the freezing regime. In contrast, for the $\rm{SYK}_{242}$ and the $\rm{SYK}_{442}$ systems, which are {\it not} non-interacting integrable, we find
\begin{equation}\label{eq:sigma242}
  \sigma=\frac{J_{LR}^2 t_R}{\Omega^3}\left\{
                                 \begin{array}{ll}
                                   \calF_{242}(J_{LR}/t_L,t_R/t_L,T_\text{eff}/t_L), & {\rm{SYK}_{242}} \\
                                   \calF_{442}(J_{LR}/J_L,t_R/J_L,T_\text{eff}/J_L), & {\rm{SYK}_{442}}.
                                 \end{array}
                               \right.
\end{equation} { Eqs.\eqref{eq:sigma222} and \eqref{eq:sigma242} may appear to be asymmetric between the two quantum dots, but this is only an artifact of our convention of choosing the scaling functions $\calF$.}
 Our numerical results support these scaling forms {(see the End Matter)}.

{\it Discussion.-} Driven many-body systems represent an emerging frontier in non-equilibrium quantum matter. Our solvable model of interacting driven quantum dots shows approximate protection from decoherence at specific drive amplitude and frequency ratios, respectively. We have analytically derived the residual decoherence rate at freezing, which is parametrically small. It is worth noting a subtlety regarding the threshold determining whether the system freezes. In our large-$N$ formalism, freezing is protected by the separation between the drive frequency and single-particle or few-particle excitation energies. There is no clear threshold for freezing in this regime. However, we expect the starting point Eq.~\eqref{eq:SD0} within the $1/\Omega$ expansion to break down when $\Omega$ becomes comparable to the spectrum of $\hat{G}_{00}$ (i.e. $\Im \hat{G}_{00}(\Omega/2)\neq 0$). Physically, this implies that $\Omega$ becomes resonant with the single-particle energy levels of $H_\text{static}$, which will lead to coherent superposition of Floquet eigenstates localized in the left and the right dots, respectively. In that regard, we can define resonance, or the breakdown of Eq.~\eqref{eq:SD0}, as the threshold for freezing.
We note in passing that when the system is driven at a high frequency, the heating is exponentially slow \cite{DAAbanin2015,TMori2016,CKuhlenkamp2020} irrespective of whether the system freezes or not. Thus the heating-rate itself is not expected to be a good diagnostic of freezing, or lack thereof.  \update{Additionally, we note that freezing occurs in both integrable and non-integrable models, as it is determined by the underlying symmetry structure of the inter-dot hopping term and the associated commutation structure with the static part of the Hamiltonian. However, the fate of the effective model at freezing can differ between these classes of models, as demonstrated here in the context of the frequency-dependent corrections to the inter-dot conductivity and the quantum chaos.} Finally, the finite-$N$ ED and large-$N$ KBE broadly agree on some of the key aspects of freezing. Investigating the influence of the clearly observable finite-$N$ fluctuations would be an interesting future direction.

We end by noting promising future directions. Holographic connections between the SYK model and gravity \cite{SSachdev2010,SSachdev2015,SSachdev2019a,AYKitaev2015,AKitaev2018b,JMaldacena2016b,UMoitra2019,LVIliesiu2021}, and the wormhole solution for coupled SYK dots \cite{JMaldacena2018,AMGarcia-Garcia2019,PGao2021,SPlugge2020,SSahoo2020,TGZhou2020,PZhang2021,PZhang2022,TGZhou2021} are well-studied. The exact solutions for dynamical freezing hint at the potential to decouple wormholes with a drive, and analyzing this within holography would be a useful exercise. Our setup can also be extended to a lattice construction. With a drive, we can decouple bipartite Hamiltonians with a globally conserved charge $N$ into two nearly decoupled sublattices with approximately conserved charges $N_A,~N_B$. 
{Furthermore, by combining this with the Yukawa-SYK model \cite{IEsterlis2019,YWang2020a,EEAldape2022,IEsterlis2021}, it is possible to investigate the fate of dynamical freezing in correlated metals.} {Both the double quantum dot and lattice setups may enable real experimental realization of dynamical freezing. Our prediction for correction to freezing, such as the inter-dot conductivity, offers a testable prospect for future experiments.} 
Studying the growth of many-body chaos and charge diffusion in space-time under a drive would shed light on the role of freezing in localizing quantum information. \update{Finally, a controlled study of dynamical freezing in a similar setup but without disorder remains an interesting future direction as well.}

{\it Acknowledgments.-} DC thanks A. Das for many insightful discussions about dynamical freezing. HG and DC are supported in part by a New Frontier Grant awarded by the College of Arts and Sciences at Cornell University and by a Sloan research fellowship from the Alfred P. Sloan foundation to DC. HG is also supported by a Wilkins postdoctoral fellowship at Cornell University. RM is supported by Fulbright-Nehru Grant No. 2877/FNDR/2023-2024 sponsored by the Bureau of Educational and Cultural Affairs of the United States Department of State.

\bibliography{SYK_auto_updated,Reference_manual}

\newpage 
\onecolumngrid
\begin{center}
    {\large \textbf{End Matter}}
\end{center}

\begin{figure}[h!]
    \centering
    \includegraphics[width=0.55\linewidth]{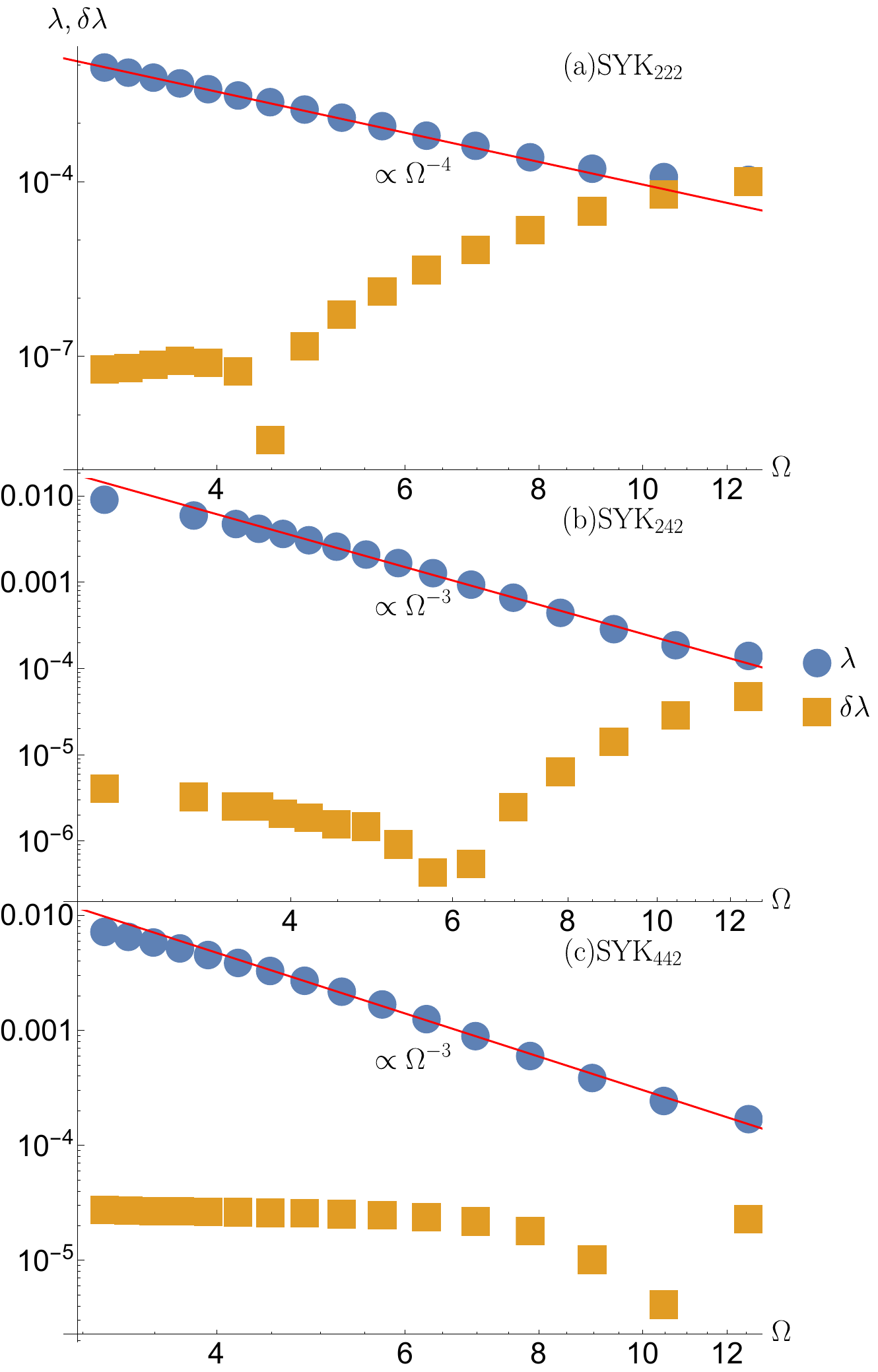}
    \caption{The decay rate $\lambda$ of $\nu_R(t)-\nu_L(t)$ at freezing as a function of driving frequency $\Omega$ for different types of coupled SYK quantum dots: (a) $\mathrm{SYK}_{222}$, (b) $\mathrm{SYK}_{242}$ and (c) $\mathrm{SYK}_{442}$. All three systems are subjected to a cosine drive with $A/\Omega$ fixed at the first freezing point. Blue circles are the decay rates $\lambda$ and yellow squares are the numerical error estimates $\delta\lambda$. The red line is a power-law fit to the $\lambda$ data points. Parameters used: (a) $t_L=t_R=0.3,t_{LR}=0.6,2\pi A/\Omega=7.555$; (b) $t_L=t_R=0.3,J_{LR}=0.6,2\pi A/\Omega=3.7775$; (c) $J_L=t_R=0.3,J_{LR}=0.6,2\pi A/\Omega=3.7775$. The step size for time evolution is fixed at $h_t=0.02$, and the system is evolved from $t=0$ upto $t=T_\text{max}=50$. }
    \label{fig:lambda}
\end{figure}

{
\textit{General criterion for freezing from the leading-order Magnus expansion} In this appendix we provide a derivation of the freezing criteria presented in Eq.\eqref{eq:criterion} of the main text.

 The ``generalized" Magnus expansion \cite{AHaldar2020}  focuses on a perturbative expansion in powers of $(1/\Omega)$ with $A/\Omega\sim O(1)$. Consider the unitary operator, $W(t)$, which transforms the Hamiltonian into the co-moving frame as,
\begin{subequations}
\beq \label{eq:theta}
  W(t) &=& \exp\bigg[-i\Theta(t)H_{\rm{drive}}\bigg],~\tn{where}~~
  \Theta(t) = \int_0^t \rd\bar{t}f(\bar{t}),\\
  H_{\tn{mov}}(t) &=& W^{\dagger}(t)\left[H(t)-i\partial_t\right]W(t) =W^\dagger(t)H_{\rm{static}} W(t).
\eeq
\end{subequations}
 {Note that $\Theta(t)$ is {\it not} related to the Heaviside function.} Clearly, the advantage is that $H_{\tn{mov}}(t) $ nominally becomes independent of the $O(H_{\tn{dynamic}})$ contribution.
We investigate the matrix elements of $H_{\tn{mov}}$ in the eigenbasis of $H_{\rm{drive}}$, $\{E_m,\ket{m}\}$,
\begin{equation}\label{}
  \braket{m|H_{\tn{mov}}(t)|n}=\exp\bigg[i\Theta(t)(E_m-E_n)\bigg]\braket{m|H_{\rm{static}}|n}\,.
\end{equation}
At leading order, the generalized Magnus expansion is given by $H^{(0)}=\frac{1}{T}\int_0^T \rd t ~H_{\tn{mov}}(t)$.
Let us now consider the two cases $\rm{I}$ and $\rm{II}$ in Eq.\eqref{eq:drives} and address the special points in phase-space where the off-diagonal matrix-elements  $\braket{m|H^{(0)}|n}$ {\it vanish} for $E_m\neq E_n$. We find $\braket{m|H^{(0)}|n} = \braket{m|H_{\rm{static}}|n}\times B_{mn}$, where
\begin{subequations}\label{eq:Bmn}
\beq
  B_{mn} &=& \frac{-2i }{A(E_m-E_n)}\left[\exp\left(i\frac{A T}{2}(E_m-E_n)\right)-1\right]~~ ({\rm Case~I}),\\
   B_{mn} &=& T J_0\left(\frac{A T}{2\pi}\left(E_m-E_n\right)\right)~~ ({\rm Case~II}).
\eeq
\end{subequations} Finally, dynamical freezing is associated with the emergent conservation where $[H^{(0)},H_\text{drive}]=0$, and therefore its location is determined by requiring that $B_{mn}$ vanishes, which yields the criterion reported in Eq.\eqref{eq:criterion}.

The leading order analysis above is only a necessary, but not sufficient, condition for dynamical freezing. First of all, the $B_{mn}$ calculated above do not take into account the part of $H_\text{static}$ that commutes with $H_\text{drive}$, which introduces an additional phase of order $(\delta E_m-\delta E_n)T$ in Eq.~\eqref{eq:Bmn}, where $\delta E_m=\braket{m|H_\text{static}|m}$. These corrections need to be small for freezing to be robust. Secondly, the phenomenon of resonance, i.e. a strong hybridization between $\ket{m},\ket{n}$, in the presence of degeneracies in the floquet quasienergy spectrum can also invalidate freezing \cite{AHaldar2020}. Both of the above issues can be overcome by working in the limit of a driving frequency larger than all pairs of $\delta E_m-\delta E_n$ with $\braket{m|H_\text{static}|n}\neq 0$, which can be achieved by $\Omega\gg \{t_\alpha,J_\alpha,V_{LR},J_{LR}\}$, i.e. the energy scale of single-particle or two-particle excitations.

 \textit{Residual charge transport at the freezing point.-} In this appendix, we provide numerical evidence that supports our analytical result of inter-dot conductivity reported in Eqs.\eqref{eq:sigma222} and \eqref{eq:sigma242}. By solving the Kadanoff-Baym equation, we obtain the densities $\nu_R(t)$ and $\nu_L(t)$ of the two dots as a function of time, respectively. The difference $\nu_R-\nu_L$ is fitted with an expoential decaying form 
 \begin{equation}
    \nu_R(t)-\nu_L(t)=A \exp(-\lambda t)\,.
 \end{equation} The decay rate $\lambda$ can be related to the conductivity $\sigma$ via the Einstein relation 
 \begin{equation}
    \lambda=\sigma\left(\frac{1}{\kappa_L}+\frac{1}{\kappa_R}\right)\,.
 \end{equation} Here $\kappa_L$ and $\kappa_R$ are the compressibilities of the two dots, respectively.  At freezing, the two dots are approximately decoupled, and we expect the compressibilities take the value of the individual isolated quantum dots, which is $\Omega$-independent.

    We compare the numerical results for $\lambda$ with our analytical results by verifying the scaling of $\lambda$ as a function of $\Omega$. We drive the three types of coupled SYK models introduced in the main text with a cosine drive and at the first freezing point, and we extract the decay rates $\lambda$ as shown in Fig.~\ref{fig:lambda}.
  To ensure that the decay rates we extract are numerically accurate, we plot it together with an error estimate $\delta \lambda$, which is extracted from the deviations from perfect conservation of the total particle density (due to numerical limitations) $\delta \lambda=3(\nu_\text{tot}(T_\text{max})-\nu_\text{tot}(0))/T_\text{max}$, where $T_\text{max}$ is the maximum propagation time of the Kadanoff-Baym equation. 
  When the drive frequency $\Omega$ is too large, our integration algorithm is incapable of tracking the rapid oscillation of the Green's functions, where the error $\delta\lambda$ grows and becomes comparable to $\lambda$. When $\Omega$ is too small so that it is comparable to the single-particle energy scales of the static Hamiltonian, the high-frequency expansion is expected to break down, so $\lambda$ will not be described by our analytical result. When $\Omega$ takes value in the intermediate regime, we observe agreement between our numerical result and our analytical expectations, as indicated by the power-law fits in Fig.~\ref{fig:lambda}.

  }

\newpage

\clearpage



\section*{Supplementary material}

**Supplementary information for "Dynamical Freezing in Exactly Solvable Models of Driven Chaotic Quantum Dots"**
Haoyu Guo, Rohit Mukherjee, Debanjan Chowdhury

### Derivation of Kadanoff-Baym Equation

In this section we provide a derivation of the Kadanoff-Baym equation for the coupled SYK quantum dots. Our starting point is the disordered averaged Keldysh path-integral

$$Z = \int Dt_L P[t_L] Dt_R P[t_R] DJ_L P[J_L] DJ_R P[J_R] DV P[V] DJ' P[J'] Dc^\dagger Dc \exp(iS).  \tag{15}$$

Here we focus on the leading large-$N$ order where only the replica symmetric saddle point is relevant, so we have set the replica number to one and replaced the quenched average by annealed average [76]. The action before averaging is

$$S = \int dz \left[ \sum_{\alpha=L,R} \sum_i c_{\alpha i}^\dagger i\partial_z c_{\alpha i} - H[c^\dagger, c] \right],  \tag{16}$$

where the coordinate $z$ runs on the Keldysh contour $C = C_\beta \cup C_+ \cup C_-$ as shown in Fig. 4, and $H[c^\dagger, c]$ is the Hamiltonian defined in the main text.

We perform the disorder averaging over the couplings, with the Gaussian integral identity

$$\langle e^{i(a\bar{z}+\bar{a}z)} \rangle = e^{-|a|^2 \langle |z|^2 \rangle},  \tag{17}$$

and we rewrite the result with $G$-$\Sigma$ action, by inserting the functional identity

$$1 = \int DG D\Sigma \exp\left[ \int dz_1 dz_2 \sum_{\alpha=L,R} \Sigma_\alpha(z_1, z_2) \left( \sum_i i c_{\alpha i}(z_2) c_{\alpha i}^\dagger(z_1) + N G_\alpha(z_2, z_1) \right) \right].  \tag{18}$$

The result is

$$Z = \int DG D\Sigma \exp\left( iNS[G, \Sigma] \right),  \tag{19}$$

where

$$\begin{aligned}
iS[G, \Sigma] = & \sum_\alpha \ln\det(i\partial_z - f_\alpha(z) - \Sigma_\alpha) + \sum_\alpha \int dz dz' \Sigma_\alpha(z, z') G_\alpha(z', z) \\
& - \int dz dz' \sum_\alpha \left[ \frac{t_\alpha^2}{2} G_\alpha(z, z') G_\alpha(z', z) + \frac{J_\alpha^2}{4} G_\alpha^2(z, z') G_\alpha(z', z)^2 \right] \\
& - \int dz dz' \left[ V_{LR}^2 G_L(z, z') G_R(z', z) + J_{LR}^2 G_L^2(z, z') G_R^2(z', z) \right].
\end{aligned}  \tag{20}$$

Here $f_\alpha(z) = s_\alpha f(t)$ when $z = t \in C_+ \cup C_-$ on the real-time branches ($s_L = 1$, $s_R = -1$), and on the imaginary branch $z = -i\tau \in C_\beta$, $f_\alpha(z) = -\mu_\alpha$ is set to the initial chemical potentials of the two dots.

Before getting to the saddle point equations, we can perform a gauge transformation on the bi-local variables

$$G_\alpha(z, z') \to e^{-i\theta_\alpha(z)} G_\alpha(z, z') e^{i\theta_\alpha(z')},  \tag{21}$$

$$\Sigma_\alpha(z, z') \to e^{-i\theta_\alpha(z)} \Sigma_\alpha(z, z') e^{i\theta_\alpha(z')}.  \tag{22}$$

where

$$\theta_\alpha(z) = \begin{cases} \int_0^z d\bar{z} f_\alpha(\bar{z}), & z \in C_+ \cup C_-; \\ 0, & z \in C_\beta. \end{cases}  \tag{23}$$

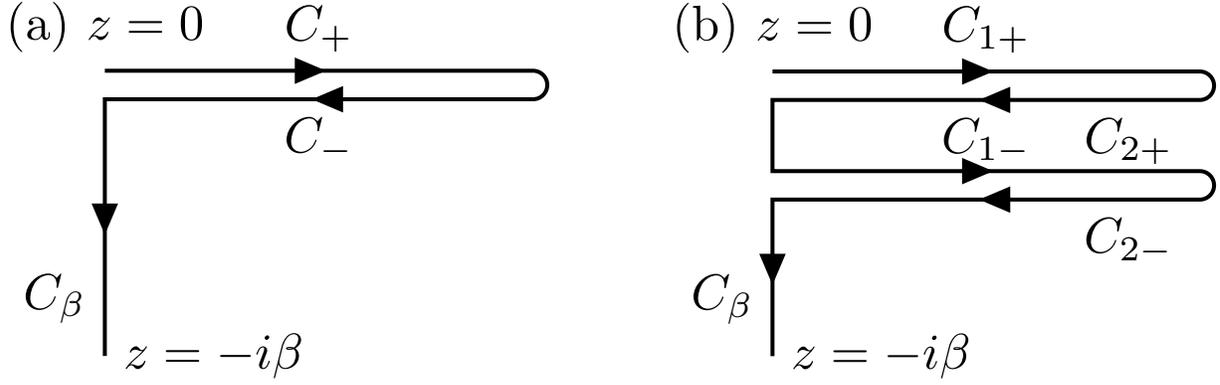

FIG. 4: (a) The Keldysh contour for calculating the time-evolution of particle density. It consists of a real-time branch $C_+ \cup C_-$ and an imaginary-time branch $C_\beta$. (b) The double-Keldysh contour for calculating the out-of-time-ordered correlator, which contains two real-time branches $C_1 = C_{1+} \cup C_{2+}$ and $C_2 = C_{2+} \cup C_{2-}$ and an imaginary-time branch $C_\beta$. The distance between $C_1$ and $C_2$ along the imaginary direction is infinitesimal and is exaggerated in the illustration.

which removes the driving term in the log determinant and transfers it into the self-energies. The action therefore reads

$$
\begin{aligned}
iS[G, \Sigma] = & \sum_\alpha \ln \det(i\partial_z - \Sigma_\alpha) + \sum_\alpha \int \mathrm{d}z \mathrm{d}z' \Sigma_\alpha(z, z') G_\alpha(z', z) \\
& - \int \mathrm{d}z \mathrm{d}z' \sum_\alpha \left[ \frac{t_\alpha^2}{2} G_\alpha(z, z') G_\alpha(z', z) + \frac{J_\alpha^2}{4} G_\alpha^2(z, z') G_\alpha(z', z)^2 \right] \\
& - \int \mathrm{d}z \mathrm{d}z' \left[ V_{LR}^2 e^{2i(\theta(z) - \theta(z'))} G_L(z, z') G_R(z', z) + J_{LR}^2 e^{4i(\theta(z) - \theta(z'))} G_L^2(z, z') G_R^2(z', z) \right].
\end{aligned}
\tag{24}
$$

At the leading large-$N$ order, we can take the saddle point of the action (24), which yields Eqs.(7),(8) in the main text.

### Solution of the Kadanoff-Baym Equation

In this section we discuss procedure of solving the Kadanoff-Baym Equation (KBE) in the main text. While the KBE is compact when written in the form of Contour Green's function $G(z_1, z_2)$ where $z_1, z_2$ are points on the complex Keldysh contour, to solve it we must expand it into components [42, 43]. These components are classified according to whether $z_1, z_2$ sits on the forward contour $C_+$, the backward contour $C_-$ or the imaginary-time contour $C_\beta$. They are

- The Matsubara component

$$
G^M(\tau_1 - \tau_2) = -iG(z_1 = -i\tau_1, z_2 = -i\tau_2),
\tag{25}
$$

where $z_1, z_2$ are both in $C_\beta$, and we assumed translation symmetry on $C_\beta$.

- The left-mixing component

$$
G^\rceil(t_1, \tau_2) = G(z_1 = t_1, z_2 = -i\tau_2),
\tag{26}
$$

where $z_1 \in C_+$ or $C_-$ and $z_2 \in C_\beta$.

- The right-mixing component

$$
G^\lceil(\tau_1, t_2) = G(z_1 = -i\tau_1, z_2 = t_2),
\tag{27}
$$

where $z_1 \in C_\beta$ and $z_2 \in C_+$ or $C_-$.

- The lesser component

$$G^<(t_1, t_2) = G(z_1 = t_1, z_2 = t_2),$$ (28)

where $z_1 \in C_+, z_2 \in C_-$.

- The greater component

$$G^>(t_1, t_2) = G(z_1 = t_1, z_2 = t_2)$$ (29)

where $z_1 \in C_-, z_2 \in C_+$.

- The regarded component

$$G^R(t_1, t_2) = \theta(t_1 - t_2) \left[ G^>(t_1, t_2) - G^<(t_1, t_2) \right].$$ (30)

Here $\theta$ is the Heaviside step function.

- The advanced component

$$G^A(t_1, t_2) = \theta(t_2 - t_1) \left[ G^<(t_1, t_2) - G^>(t_1, t_2) \right].$$ (31)

The self-energies can be decomposed in a similar way.

Since the Green's function is anti hermitian $G(z_2, z_1)^* = -G(z_1^*, z_2^*)$, knowing the 4 components $(G^M, G^\rceil, G^<, G^R)$ determines the full $G(z_1, z_2)$. The Kadanoff-Baym equation in component form reads [43] (we suppress the $L, R$ indices for simplicity):

$$-\partial_\tau G^M(\tau) - \int_0^\beta d\tau' \Sigma^M(\tau - \tau') G^M(\tau') = \delta(\tau),$$ (32)

$$i\partial_t G^\rceil(t, \tau) - \int_0^t d\bar{t}\, \Sigma^R(t, \bar{t}) G^\rceil(\bar{t}, \tau) = \int_0^\beta d\tau' \Sigma^\rceil(t, \tau') G^M(\tau' - \tau),$$ (33)

$$i\partial_t G^R(t, t') - \int_{t'}^t d\bar{t}\, \Sigma^R(t, \bar{t}) G^R(\bar{t}, t') = \delta(t - t'),$$ (34)

$$i\partial_t G^<(t, t') - \int_0^t d\bar{t}\, \Sigma^R(t, \bar{t}) G^<(\bar{t}, t') = \int_0^{t'} d\bar{t} \Sigma^<(t, \bar{t}) G^A(\bar{t}, t') - i \int_0^\beta d\tau \Sigma^\rceil(t, \tau) G^\lceil(\tau, t').$$ (35)

We use the NESSi package [43] to solve Eqs.(32)-(35). We first solve Eq.(32) on the Matsubara axis by fixed point iteration. Here we discretize the interval $[0, \beta]$ by $N_\tau = 400$ segments of equal length. Next, we use the Matsubara solution as initial condition and propagate Eqs.(33)-(35) on the real-time axis. The real-time coordinates are discretized in timestep $h_t$. We typically use $h_t = 0.025T$. The propagation utilizes a simple predictor-corrector algorithm. We start from an initial guess of $G(t, t')$ based on 5th order polynomial extrapolation of $G(t - n h_t, t'), n = 1, 2, 3, 4, 5$. From this guess we update the self-energies, and then solve the discretized versions of Eqs.(33)-(35) to update $G(t, t')$, and we iterate until $G(t, t')$ converges (usually within 2 or 3 iterations). The derivatives in Eqs.(33)-(35) are discretized using 5th order backward differentiation, and the convolutions are calculated with 5th order Gregory quadratures. Therefore, we expect our numerical solution to have 5th order accuracy in $h_t$.

### Computation of quantum chaos

In this section, we discuss the computation of the inter-branch correlation function $G^{12}(z_1, z_2)$, which serves as a proxy for OTOC and quantum chaos. As discussed in Ref.[46], it can be viewed as the response of the system to perturbation of the thermofield double state, which can be described by the Kadanoff-Baym equation on the double-Keldysh contour (see Fig.4) $C = C_1 \cup C_2 \cup C_\beta$. Unlike most theoretical literatures where $C_1$ and $C_2$ are separated by distance $\beta/2$ in the imaginary direction, here they are infinitesimally next to each other. As we will see later this configuration allows us to reuse the numerical solution from the single Keldysh-contour. The perturbation can be characterized as an additional term in the self-energy:

$$\delta\Sigma_\alpha(t_1, t_2) = \delta(t_1)\delta(t_2) \times \begin{cases} -iz, & \text{if } t_1 \in C_1, t_2 \in C_2, \\ iz, & \text{if } t_2 \in C_1, t_1 \in C_2. \end{cases}$$ (36)

Note here that $\delta\Sigma_\alpha(t_1, t_2)$ exists for both $t_1 \in C_{1+}$ and $t_1 \in C_{1-}$ and it takes equal value, and similar statements holds for $t_2$ as well. This is required to ensure that forward propagation $\int_{C_{1+}} \mathrm{d}z$ and backward propagation $\int_{C_{1-}} \mathrm{d}z$ cancel exactly.

We analyze the Kadanoff-Baym equation in component form. For simplicity we suppress the $\alpha = L, R$ indices. In contour-coordinate form, the first KB equation reads

$$i\partial_z G(z, z') - \int_C \mathrm{d}\bar{z}\, \Sigma(z, \bar{z}) G(\bar{z}, z') = \delta(z, z')\,. \tag{37}$$

We first consider the case where $z, z' \in C_\beta \cup C_1$, which corresponds to the Green's function components introduced in the previous section. The difference between the single Keldysh contour case is that the convolution variable $\bar{z}$ can sit on $C_2$. Since $\bar{z}$ does not coincide with $z, z'$ as they are in different branches, the integrand $\Sigma(z, \bar{z}) G(\bar{z}, z')$ is continuous. Furthermore, it takes equal value for $\bar{z} \in C_{2+}$ and $\bar{z} C_{2-}$, and therefore $\int_{C_2} \mathrm{d}\bar{z}\, \Sigma(z, \bar{z}) G(z, z') = 0$ because the time-arrowed is reversed from $C_{2+}$ to $C_{2-}$. The bottom line is that the inter-branch components $G^M, G^R, G^<$ and the mixing components $G^\rceil$ are the same as the single-Keldysh contour case. Similar discussion also holds for $z, z' \in C_\beta \cup C_2$.

Next, we restrict $z = t_1 \in C_1$, $z' = t_2 \in C_2$ in Eq.(7) and break up the convolution in different components. We find that the equation for $G^{12}(t_1, t_2)$ is identical to Eq.(35):

$$i\partial_t G^{12}(t, t') - \int_0^t \mathrm{d}\bar{t}\, \Sigma^R(t, \bar{t}) G^{12}(\bar{t}, t') = \int_0^{t'} \mathrm{d}\bar{t}\, \Sigma^{12}(t, \bar{t}) G^A(\bar{t}, t') - i \int_0^\beta \mathrm{d}\tau \Sigma^\rceil(t, \tau) G^\lceil(\tau, t')\,. \tag{38}$$

The conjugate equation is

$$-i\partial_{t'} G^{12}(t, t') - \int_0^{t'} \mathrm{d}\bar{t}\, G^{12}(t, \bar{t}) \Sigma^A(\bar{t}, t') = \int_0^t \mathrm{d}\bar{t}\, G^R(t, \bar{t}) \Sigma^{12}(\bar{t}, t') - i \int_0^\beta \mathrm{d}\tau G^\lceil(t, \tau) \Sigma^\rceil(\tau, t')\,. \tag{39}$$

Here the self-energies are calculated from Eq.(8) of the main text plus the perturbation (36).

Similarly, for $G^{21}(t_1, t_2)$ where $z = t_1 \in C_2$, $z' = t_2 \in C_1$, the resulting equation is of the same form

$$i\partial_t G^{21}(t, t') - \int_0^t \mathrm{d}\bar{t}\, \Sigma^R(t, \bar{t}) G^{21}(\bar{t}, t') = \int_0^{t'} \mathrm{d}\bar{t}\, \Sigma^{21}(t, \bar{t}) G^A(\bar{t}, t') - i \int_0^\beta \mathrm{d}\tau \Sigma^\rceil(t, \tau) G^\lceil(\tau, t')\,, \tag{40}$$

$$-i\partial_{t'} G^{21}(t, t') - \int_0^{t'} \mathrm{d}\bar{t}\, G^{21}(t, \bar{t}) \Sigma^A(\bar{t}, t') = \int_0^t \mathrm{d}\bar{t}\, G^R(t, \bar{t}) \Sigma^{21}(\bar{t}, t') - i \int_0^\beta \mathrm{d}\tau G^\lceil(t, \tau) \Sigma^\rceil(\tau, t')\,. \tag{41}$$

Therefore, one particular solution of Eqs.(38)-(41) is $G^{12} = G^<$ and $G^{21} = G^>$. However, unlike the intra-branch component $G^>, G^<$ which satisfy the relation $G^> - G^< = G^R - G^A$, $G^{12}$ and $G^{21}$ are not constrained by this relation, so they are susceptible to perturbations, and can flow to the solution $G^{12} = G^{21} = 0$ at $t \to \infty$.

Our goal is therefore to solve Eqs.(38)-(41) numerically. First, we cannot literally use the delta-function perturbation (36), but instead we integrate it to transform it into an initial condition:

$$G^{12}(t, t') = G^<(t, t') + z\left[G^>(t, t') - G^<(t, t')\right]\,, t > 0, t' = 0 \text{ or } t = 0, t' > 0\,. \tag{42}$$

$$G^{21}(t, t') = G^>(t, t') - z\left[G^>(t, t') - G^<(t, t')\right]\,, t > 0, t' = 0 \text{ or } t = 0, t' > 0\,. \tag{43}$$

We then propagate Eqs.(38)-(41) in the first quadrant of $(t, t')$-plane with initial condition Eqs.(42)-(43). The propagation algorithm is the same as that for the single Keldysh contour KB equation. Because the NESSi package only supports Green's function on a single Keldysh contour, we have implemented our own solver in Julia.

Finally, we benchmark our solver using the single-dot complex SYK model, whose self-energy is

$$\Sigma(z, z') = J^2 G^2(z, z') G(z', z)\,. \tag{44}$$

We compare our result with the direct solution of the Schwinger-Dyson equation on the double-Keldysh contour [46]. We discretize the double Keldysh contour $C$ into $L$ segments, indexed by $m = 1, 2, \ldots, L$. The contour ordered Green's function is then discretized as a matrix $\hat{G}(m, n) = G(mh_t, nh_t)$ with $h_t$ being the step size. The discretized Schwinger-dyson equation reads

$$\hat{G} = \frac{1}{\hat{G}_0^{-1} - \hat{\Sigma}}\,, \tag{45}$$

where $\hat{G}_0$ is the free fermion Green's function, and $\hat{\Sigma}$ is the discretized fermion self-energy. It is given by

$$\hat{\Sigma}(m,n) = h_t^2 J^2 \hat{G}(m,n)^2 \hat{G}(n,m) f(m) f(n) - iz\sigma(m,n).$$  (46)

Here $f(m) = 1, -1, -i$ indicates the direction of contour integration. The source term $\sigma(m,n)$ is ($\delta(\cdot,\cdot)$ below is Kronecker delta)

$$\sigma(m,n) = \delta(m,n_0(C_{1+}))\delta(n,n_0(C_{2+})) + \delta(m,n_0(C_{1-}))\delta(n,n_0(C_{2-})) - \delta(m,n_0(C_{1+}))\delta(n,n_0(C_{2-})) - \delta(m,n_0(C_{1-}))\delta(n,n_0(C_{2+})) - (m \leftrightarrow n),$$  (47)

where $n_0(C)$ is the index of the point $t = 0$ on branch $C$.

The comparison is shown in Fig. 5. The only noticeable difference between our method and the benchmark is at initial times where the initial conditions are implemented differently.

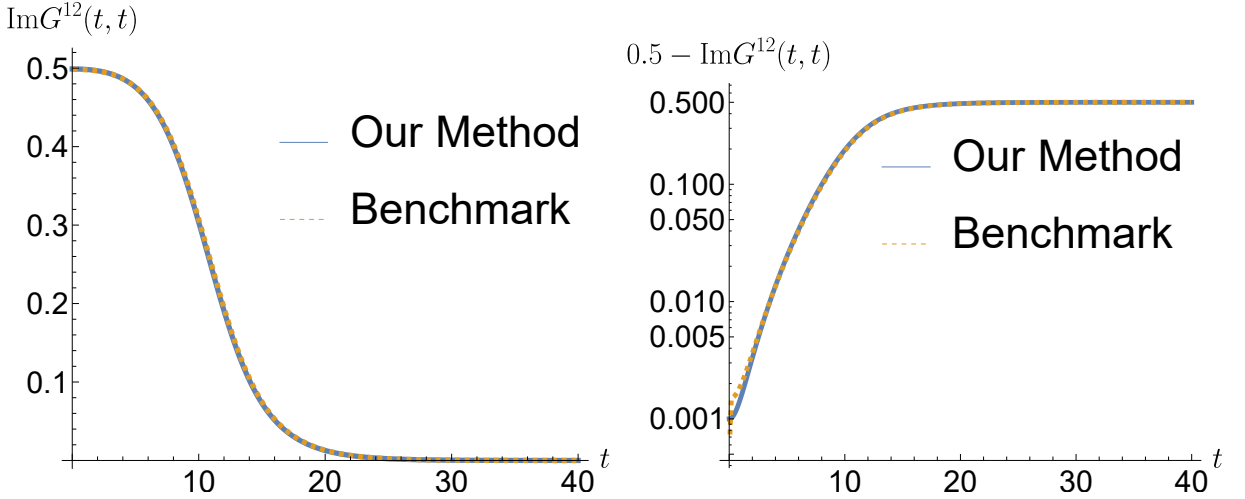

FIG. 5: Comparing our chaos solver with Ref. [46]. We consider the complex SYK model at half filling, with $J = 1$, $\beta J = 8$ and $z = 0.001$. In our algorithm we used time step $h_t = 0.04$ and in the benchmark we used $h_t = 0.125$. The left panel compares $\text{Im}G^{12}(t,t)$ in linear scale, and the right panel compares $0.5 - \text{Im}G^{12}(t,t)$ in log scale. The only noticeable difference is at the initial times which is due to different implementation of the initial condition.

### ED results with disorder average

In Fig.(6) and Fig.(7) we present the exact diagonalization results after averaging over different disorder realizations. We observe that the qualitative aspects are very similar to the single-snapshot data presented in the main text.

### Freezing for different initial conditions

In this section, we present additional data which demonstrates freezing at different initial conditions. In Fig. 8, we show large-$N$ results where we initialize the system at different initial temperatures $\beta^{-1}$ with a chemical potential difference $\mu_R - \mu_L$, and we see freezing holds irrespective of the temperature. We note that even for the high initial temperature state ($\beta/T = 1$) where $\beta^{-1} \gg t_\alpha, J_\alpha, V_{LR}, J_{LR}$, freezing is still present. We note by construction of the KBE, the initial states considered in large-$N$ are the thermal density matrices.

In Fig. 9 we show results obtained from ED. Here instead of the thermal density matrix we consider the thermal pure state

$$|\psi_\beta\rangle = \sum_n e^{-\beta E_n} |n\rangle,$$  (48)

where $n$ samples over states with a given initial $N_R - N_L$.

As shown in Fig. 9 states of different initial temperature $\beta^{-1}$ freezes at the theoretical predicted freezing points.

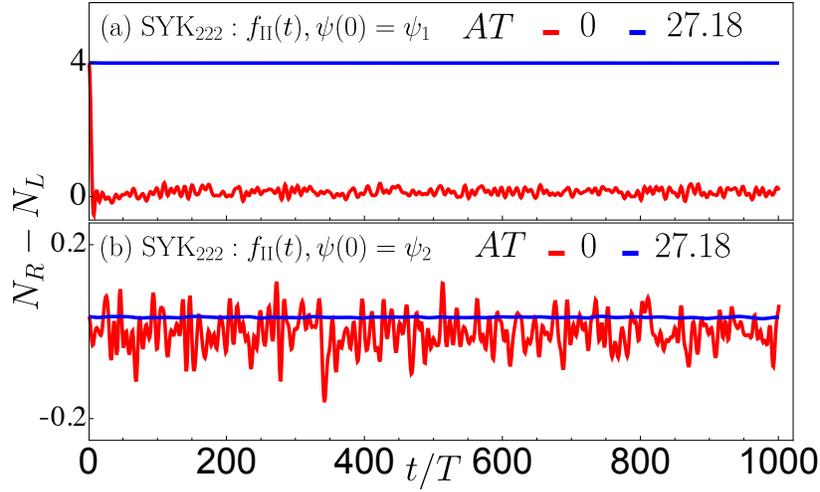

FIG. 6: ED results averaged over fifty disorder realizations for the difference in particle numbers of two dots, $N_R - N_L$, plotted stroboscopically. (a) Freezing dynamics yields results identical to those from individual snapshots. The fluctuations around the thermalized value of the difference are suppressed when we are away from freezing. (b) A similar feature is observed when the system is initially prepared in a state such that the difference of $N_R - N_L$ is very small. All other parameters for the plot is same as Fig(1)

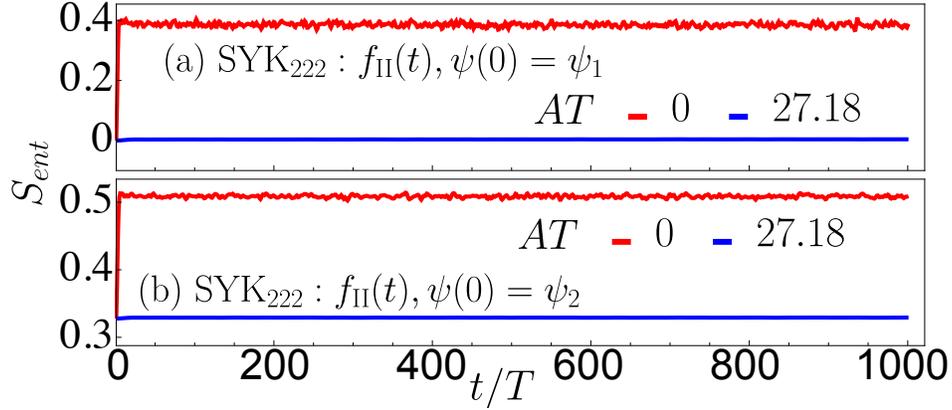

FIG. 7: Disorder averaged (fifty realizations) Entanglement entropy ($S_{ent}$) is plotted stroboscopically at and away from the freezing point. (a) Results are mirrored by the outcomes of individual snapshots as in Fig.(2). (b) Initial state that corresponds to small difference in particle number has nonzero entanglement entropy to begin with. All other parameters for the plot is same as Fig(1)

### Analytical Estimate for the density decay rate at freezing

In this section we provide more details for the analytical estimate of the leading correction to the freezing behavior. The goal of this section is to calculate the inter-dot conductance $\sigma$, which serves as an estimate for the decay rate of the density difference $N_R - N_L$. For simplicity, we will restrict to the linear response regime near the Floquet steady state.

### Green's function in the Floquet steady state

At the Floquet steady state, the Green's functions possess discrete time translation symmetry

$$G(t, t') = G(t + T, t' + T).$$  (49)

The fourier frequency $\omega, \omega'$ corresponding to $t, t'$ must differ by integral multiples of $\Omega = 2\pi/T$, and we therefore consider the following Floquet fourier transform

$$\hat{G}_{mn}(\omega) = \frac{1}{T} \int_0^T \mathrm{d}t \int_{-\infty}^{\infty} \mathrm{d}t' G(t, t') e^{i(\omega + m\Omega)t} e^{-i(\omega + n\Omega)t'}.$$  (50)

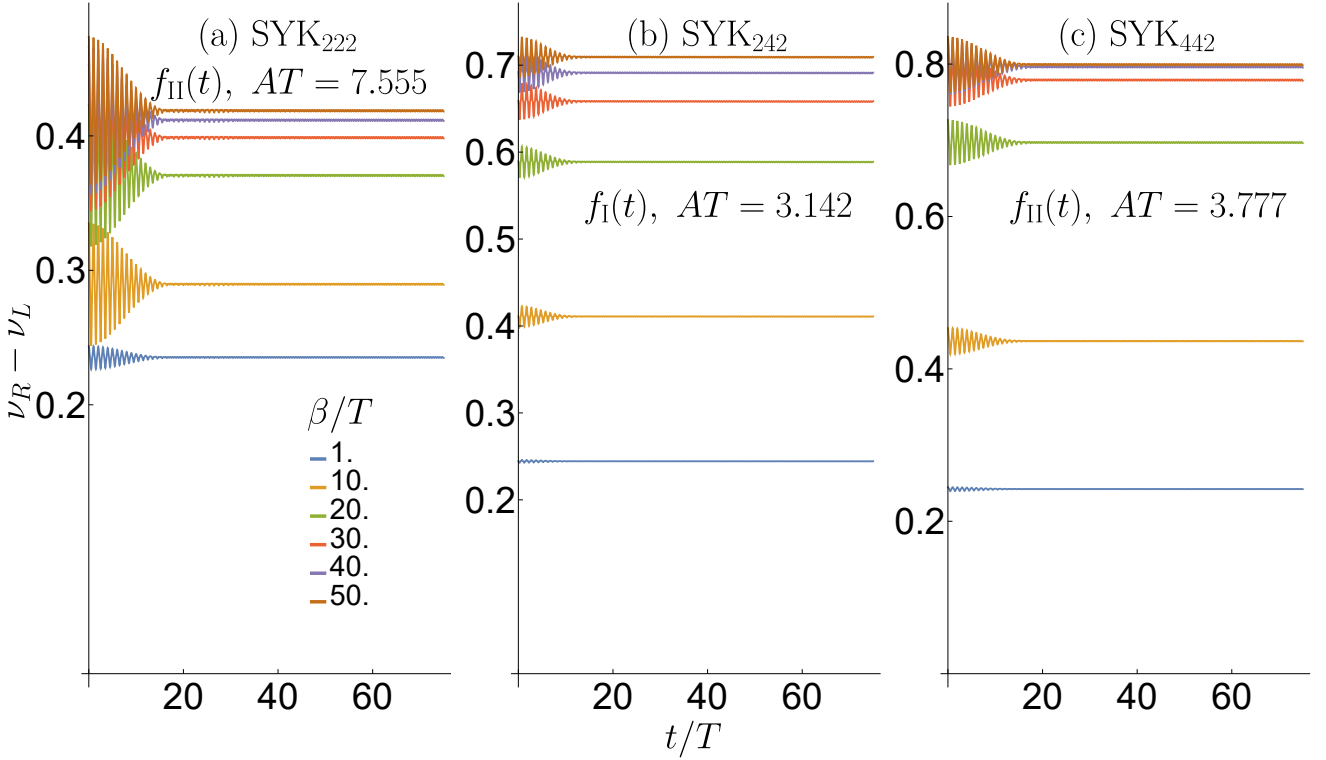

FIG. 8: Freezing in the large-N limit with different initial temperatures ($\beta^{-1}$). (a) SYK$_{222}$ model with cosine drive at the first freezing point; (b) SYK$_{242}$ model with square wave drive at the first freezing point; (c) SYK$_{442}$ model with cosine drive at the frist freezing point; Parameters used: $t_L T = J_L T = t_R T = J_R T = (1/2)V_{LR}T = (1/2)J_{LR}T = 0.1$, We use chemical potential $\mu_L T = -\mu_R T = 0.1$ for initial inverse temperature $\beta/T = 10, 20, 30, 40, 50$, and we use $\mu_L T = \mu_R T = 0.5$ for the initial inverse temperature $\beta/T = 1$.

This definition has a redundancy $\hat{G}_{mn}(\omega + \Omega) = \hat{G}_{m+1,n+1}(\omega)$, which can be removed by restricting $-\Omega/2 < \omega \leq \Omega/2$. Therefore $\omega$ can be interpreted as the Floquet quasi-energy. The inverse transform is

$$G(t, t') = \sum_{mn} \int_{-\Omega/2}^{\Omega/2} \frac{\mathrm{d}\omega}{2\pi} \hat{G}_{mn}(\omega) e^{-i(\omega + m\Omega)t} e^{i(\omega + n\Omega)t'} \,. \tag{51}$$

It can be verified that for convolution $C(t, t') = \int \mathrm{d}\bar{t} A(t, \bar{t}) B(\bar{t}, t')$, the corresponding Floquet fourier transform becomes matrix multiplication

$$\hat{C}(\omega) = \hat{A}(\omega)\hat{B}(\omega) \,. \tag{52}$$

For point-wise multiplication $C^+(t, t') = A(t, t')B(t, t')$ and $C^-(t, t') = A(t, t')B(t', t)$, the Floquet fourier transform becomes

$$\hat{C}_{mn}^+(\omega) = \sum_{m_1, n_1} \int_{-\Omega/2}^{\Omega/2} \frac{\mathrm{d}\omega_1}{2\pi} \hat{A}_{m_1 n_1}(\omega_1) \hat{B}_{m-m_1, n-n_1}(\omega - \omega_1) \,, \tag{53}$$

$$\hat{C}_{mn}^-(\omega) = \sum_{m_2, n_2} \int_{-\Omega/2}^{\Omega/2} \frac{\mathrm{d}\omega_2}{2\pi} \hat{A}_{m_2+m, n_2+n}(\omega + \omega_2) \hat{B}_{m_2 n_2}(\omega_2) \,. \tag{54}$$

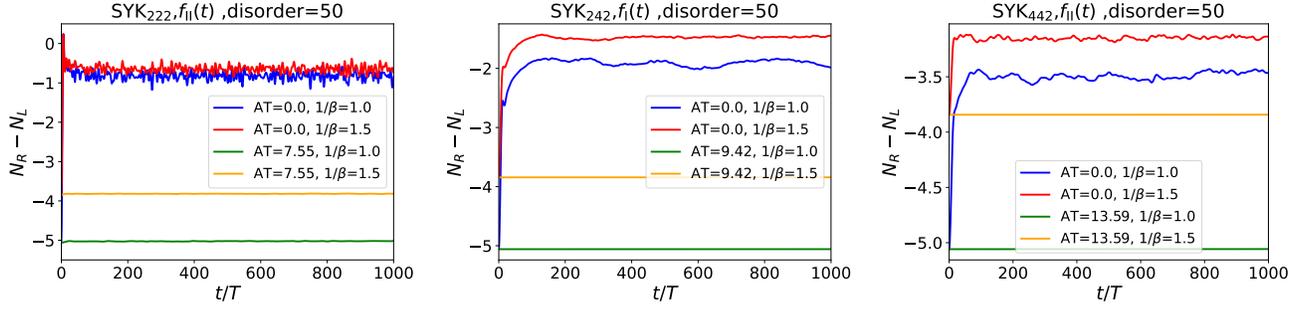

FIG. 9: ED results for the freezing at different initial temperature: We use chemical potential $\mu_L = -\mu_R = 0.055$; For SYK$_{222}$: $t_L T = t_R T = (1/2)V_{LR}T = 0.1$ under drive (2b), SYK$_{242}$ with $t_L T = t_R T = (1/2)J_{LR}T = 0.1$ under drive (2a) and SYK$_{442}$ with $J_L T = t_R T = J_{LR} T = 0.1$ under drive (2b). $\beta$ is the inverse tememprature with $k_B = 1$.

Next we discuss the structure of the Kadanoff-Baym or Schwinger-Dyson equation at the steady state. The correlation with the initial state has died out, so we can drop the $\lceil$ and $\rceil$ components in the Kadanoff-Baym equations. The retarded Green's function is given by the equation (here we use the gauge transformed Green's function introduced in (21))

$$i\partial_t G^R(t,t') - \int_{t'}^{t} d\tilde{t} \Sigma^R(t,\tilde{t}) G^R(\tilde{t},t') = \delta(t,t'), \tag{55}$$

and the lesser Green's function satisfies

$$i\partial_t G^<(t,t') - \int_{-\infty}^{t} d\tilde{t} \Sigma^R(t,\tilde{t}) G^<(\tilde{t},t') = \int_{-\infty}^{t'} d\tilde{t} \Sigma^<(t,\tilde{t}) G^A(\tilde{t},t'). \tag{56}$$

After Floquet fourier transform, Eq.(55) becomes

$$(\omega + \hat{m}\Omega - \hat{\Sigma}^R(\omega))\hat{G}^R(\omega) = \hat{1}. \tag{57}$$

Here $\hat{m}_{ab} = a\delta_{ab}$.

The self-energy is given by Eq. 8 of the main text, which we rewrite by separating out the static intra-dot contribution and the oscillating inter-dot contribution

$$\Sigma_\alpha^R(t,t') = \Sigma_\alpha^{R,a}(t,t') + e^{is_\alpha\varphi(t)}\Sigma_\alpha^{R,b}(t,t') e^{-is_\alpha\varphi(t')}. \tag{58}$$

Here $\varphi = 2\theta$ or $4\theta$ where $\theta$ is given by Eq. 3a of the main text, depending on which inter-dot interaction is used. We have restored the dot index $\alpha = L, R$ here with $s_L = -s_R = 1$. The Floquet fourier transform of (58) reads

$$\hat{\Sigma}_\alpha^R = \hat{\Sigma}_\alpha^{R,a} + \hat{F}_\alpha \hat{\Sigma}_\alpha^{R,b} \hat{F}_\alpha^\dagger. \tag{59}$$

Here $\hat{F}_L = \hat{F}_R^\dagger = \hat{F}$. The unitary matrix $\hat{F}_{mn} = F_{m-n}$ is defined by the fourier coefficients of $e^{i\varphi}$:

$$e^{i\varphi(t)} = \sum_m F_m e^{-im\Omega t}. \tag{60}$$

### Kubo formula for inter-dot conductance

To estimate the approximate conservation tied to the decoupled charge densities, we calculate the inter-dot conductivity. In this subsection we derive a Kubo formula for the conductance. Within linear response, the current flowing into the left dot is given by

$$J(t) = -\int dt' \Pi_{J,n_L}^R(t,t')(V_R(t') - V_L(t')). \tag{61}$$

Here $\Pi_{J,n_L}^R$ is the retarded correlation function between $J$ and $n_L$, and $V_R$ and $V_L$ are the electrostatic potential on the two dots respectively. Since $J = \dot{n}_L$, we have

$$\partial_t \Pi_{J,n_L}^R(t,t') = \text{contact term} + \Pi_{J,J}^R(t,t'). \tag{62}$$

The inter-dot conductance can therefore be written as

$$\sigma(t,t') = -\Pi^R_{J,n_L}(t,t') = \sum_{m,n} \int_{-\Omega/2}^{\Omega/2} \frac{d\omega}{2\pi} \frac{\Pi^R_{JJ,mn}(\omega) - \Pi^R_{JJ,mn}(0)}{-i(\omega + n\Omega)} e^{-i(\omega+m\Omega)t} e^{i(\omega+n\Omega)t'} \; . \tag{63}$$

Since we are interested in the slow dynamics, with time scale much longer than one period $T$, we consider the averaged conductance over one-period:

$$\sigma(\omega) = \frac{1}{T} \int_0^T d\tau \int_{-\infty}^{\infty} dt \sigma(\tau + t/2, \tau - t/2) e^{i\omega t} = \frac{\Pi^R_{JJ,00}(\omega) - \Pi^R_{JJ,00}(0)}{-i\omega} \; . \tag{64}$$

The real part can be rewritten as

$$\mathrm{Re}\,\sigma(\omega) = \frac{i}{2\omega} \left[ \Pi^R_{00}(\omega) - \Pi^A_{00}(\omega) \right] \; , \tag{65}$$

where we have dropped the $JJ$ subscript for clarity, and from now on $\Pi^{R/A}$ will refer to the retarded/advanced current-current correlator. We are interested in the DC limit $\omega \to 0$.

The explicit expression for the current-current correlator in the leading large-$N$ order is

$$\Pi^\gtrless(t,t')/N = -iV_{LR}^2 \left[ e^{2i\theta(t)} G_R^\gtrless(t,t') G_L^\lessgtr(t',t) e^{-2i\theta(t')} + e^{-2i\theta(t)} G_L^\gtrless(t,t') G_R^\lessgtr(t',t) e^{i\theta(t')} \right]$$
$$- 2iJ_{LR}^2 \left[ e^{4i\theta(t)} G_R^\gtrless(t,t')^2 G_L^\lessgtr(t',t)^2 e^{-4i\theta(t')} + e^{-4i\theta(t)} G_L^\gtrless(t,t')^2 G_R^\lessgtr(t',t)^2 e^{4i\theta(t')} \right] \; . \tag{66}$$

To evaluate Eq.(65), we use the identity $\Pi^R - \Pi^A = \Pi^> - \Pi^<$, and we rewrite the Green's functions using the combination $A_\alpha = i(G_\alpha^> - G_\alpha^<)$ and $G_\alpha^<$, we obtain

$$\Pi^>(t,t') - \Pi^<(t,t') = e^{2i\theta(t)} P_{RL}(t,t') e^{-2i\theta(t')} + e^{-2i\theta(t)} P_{LR}(t,t') e^{2i\theta(t')}$$
$$+ e^{4i\theta(t)} Q_{RL}(t,t') e^{-4i\theta(t')} + e^{-4i\theta(t)} Q_{LR}(t,t') e^{4i\theta(t')} \; , \tag{67}$$

where

$$P_{\alpha\beta}(t,t') = NV_{LR}^2 \left[ G_\alpha^<(t,t') A_\beta(t',t) - A_\alpha(t,t') G_\beta^<(t',t) \right] \; , \tag{68}$$

$$Q_{\alpha\beta}(t,t') = -iNJ_{LR}^2 \left[ G_\alpha^<(t,t')^2 A_\beta(t',t)^2 - A_\alpha(t,t')^2 G_\beta^<(t',t)^2 + 2iG_\alpha^<(t,t')^2 A_\beta(t',t) G_\beta^<(t',t) - 2iA_\alpha(t,t') G_\alpha^<(t,t') G_\beta^<(t',t)^2 \right] \; . \tag{69}$$

In what follows we only consider the case with a single-type of inter-dot interaction, setting either $V_{LR} = 0$ or $J_{LR} = 0$.

### High frequency expansion

The discussion so far is general and exact. Starting from this subsection we will apply the high-frequency expansion to the problem. The starting point of the expansion is to assume that the driving frequency $\Omega$ is much larger than any other single-particle energy scale of the problem (see the discussion in the main text). Therefore, the $00$−th block of the Green's function $\hat{G}$ and self-energy $\hat{\Sigma}$ are much larger than the other blocks. The leading order Schwinger-Dyson equation becomes

$$(\omega - \hat{\Sigma}_{00}^R)\hat{G}_{00}^R(\omega) = 1 \; , \tag{70}$$

and the self-energy can also be evaluated by just retaining the 00th blocks

$$\hat{\Sigma}_{\alpha,00}^R = \hat{\Sigma}_{\alpha,00}^{R,a}[\hat{G}_{00}] + |F_0|^2 \hat{\Sigma}_\alpha^{R,b}[\hat{G}_{00}] \; . \tag{71}$$

Strictly speaking, there should be another set of equations for the lesser components $\hat{G}^<, \hat{\Sigma}^<$ which are to be solved self-consistently in conjunction with Eqs.(70),(71). However, here we employ the generalized Kadanoff-Baym ansatz [77, 78], where

$$\hat{G}_{\alpha,mn}^<(\omega) = i\hat{A}_{\alpha,mn}(\omega) n_F(\omega) \; , \tag{72}$$

where $\hat{A} = i(\hat{G}^R - \hat{G}^A)$ is the spectral weight. We further assume that $n_F(\omega)$ resembles a thermal distribution at low-energies:

$$n_F(\omega) = \frac{1}{e^{\beta_{\text{eff}}\omega} + 1},\tag{73}$$

with an inverse effective temperature $\beta_{\text{eff}}$. This behavior of $G^<$ has been verified numerically in previous studies of driven SYK models [35].

To summarize, Eqs.(70)-(72) form the leading order result of the $1/\Omega$ expansion. Note that they are very similar to a non-driven system, except that the frequency variable $\omega$ is restricted to $[-\Omega/2, \Omega/2]$ in the driven setting, which is not important in the large $\Omega$ limit. The result here aligns with the spirit of the Magnus expansion where the leading order Floquet Hamiltonian can be replaced by the one-period average of the driven Hamiltonian.

For generic driving conditions, $F_0 \neq 0$ and Eq.(71) describes a coupled quantum dot. However, exactly at the freezing points discussed in the main text, $F_0$ vanishes and we obtain two decoupled quantum dots at the leading order.

### Inter-dot conductance away from freezing point

As a simpler example, let us look at the inter-dot conductance away from freezing. First we need to understand what terms enter into the DC conductance. We consider the Floquet fourier transform of (67)

$$\text{Re}\,\sigma(\omega) = \frac{i}{2\omega} \sum_{m,n} \left[ F_{-m} F_{-n}^* \hat{S}_{LR,mn}(\omega) + F_m^* F_n \hat{S}_{RL,nm}(\omega) \right].\tag{74}$$

Here $\hat{S} = \hat{P}$ or $\hat{Q}$, depending on which type of inter-dot coupling is considered. Fourier transforming Eqs.(68) and (69), we obtain

$$\frac{i}{2\omega} \hat{P}_{\alpha\beta,mn}(\omega) = N\frac{V_{LR}^2}{2} \sum_{m_2,n_2} \int_{-\Omega/2}^{\Omega/2} \frac{d\omega_2}{2\pi} \hat{A}_{\alpha,m-m_2,n-n_2}(\omega_2 + \omega)\hat{A}_{\beta,m_2 n_2}(\omega_2) \frac{n_F(\omega_2) - n_F(\omega + \omega_2)}{\omega}.\tag{75}$$

$$\begin{aligned}
\frac{i}{2\omega} \hat{Q}_{\alpha\beta,mn}(\omega) = {}& N\frac{J_{LR}^2}{2} \frac{\sinh(\beta_{\text{eff}}\omega/2)}{\omega} \sum_{m_2,n_2,m_3,n_3,m_4,n_4} \int_{-\Omega/2}^{\Omega/2} \frac{d\omega_2 d\omega_3 d\omega_4}{(2\pi)^3} \\
& \times \frac{1}{8\cosh(\beta_{\text{eff}}\omega_2/2)\cosh(\beta_{\text{eff}}\omega_3/2)\cosh(\beta_{\text{eff}}\omega_4/2)\cosh(\beta_{\text{eff}}(\omega + \omega_3 + \omega_4 - \omega_2)/2)} \\
& \times \hat{A}_{\alpha,m+m_3+m_4-m_2,n+n_3+n_4-n_2}(\omega + \omega_3 + \omega_4 - \omega_2)\hat{A}_{\alpha,m_2 n_2}(\omega_2)\hat{A}_{\beta,m_3 n_3}(\omega_3)\hat{A}_{\beta,m_4 n_4}(\omega_4).
\end{aligned}\tag{76}$$

At a non-freezing point, the largest term of Eq.(74) is from $m = n = 0$, and the largest contribution of Eqs.(75) and (76) also arise from the 00-th block of the spectral weights. We therefore conclude that $\text{Re}\,\sigma(\omega)$ is of order $1/\Omega^0$ away from freezing, implying that the system equilibrates to $N_R = N_L$ fast.

To understand how the conductance depends on other model parameters, we need the solution of Eqs.(70)-(72), and the conductance can be inferred from Eqs.(75) or (76). Some examples include

- For the SYK$_{222}$ model with $t_L = t_R = t$, there is a simple solution

$$\hat{A}_{\alpha,00}(\omega) = \frac{\sqrt{4h^2 - \omega^2}}{h^2}, \quad h = \sqrt{t^2 + |F_0|^2 V_{LR}^2}.\tag{77}$$

  The DC conductance can be estimated by power counting Eq.(75), with the result

$$\text{Re}\,\sigma(0) \sim N\frac{|F_0|^2 V_{LR}^2}{t^2 + |F_0|^2 V_{LR}^2}.\tag{78}$$

  Note here that $F_0$ depends on the ratio between driving amplitude and frequency $A/\Omega$.

- For the SYK$_{242}$ model with $t_L = t_R = t$ and at weak coupling $J_{LR} \ll t$ and low effective temperature, the spectral weight $\hat{A}_{00}(\omega)$ is similar to an isolated SYK$_2$, and the conductance is proportional to $T_{\text{eff}}^2$:

$$\text{Re}\,\sigma(0) \sim N|F_0|^2 |J_{LR}|^2 \frac{T_{\text{eff}}^2}{t^4}.\tag{79}$$

- Consider SYK$_{442}$ at weak coupling $J_{LR} \ll J_L, t_R$. The $R$-dot is a SYK$_2$ with $\hat{A}_{00}(\omega) \sim 1/t_R$, and the $L$-dot is a SYK$_4$ with $\hat{A}_{00}(\omega) \sim 1/\sqrt{J_L \omega}$. The conductance is then

$$\text{Re}\,\sigma(0) \sim N|F_0|^2 J_{LR}^2 \frac{T_{\text{eff}}}{J_L t_R^2}\,. \tag{80}$$

**Inter-dot conductance at freezing**

Now we shift our focus to the freezing point where $F_0 = 0$. Clearly the leading order estimates above all vanish and we need to go to higher order in $1/\Omega$. The main contribution to the conductance (74) now comes from terms with $m, n \neq 0$.

The task is therefore to find the leading order contribution to Eqs.(75) and (76) with $m, n \neq 0$. Heuristically, because $\hat{A}_{00}$ is the largest one, we would like to have as many zeroes in the floquet band indices $m_2, n_2, \ldots$ as possible. This leads us to consider terms of the form $\hat{A}_{mn}\hat{A}_{00}$ or $\hat{A}_{m0}\hat{A}_{0n}$ in Eq.(75), and terms of the form $\hat{A}_{mn}\hat{A}_{00}\hat{A}_{00}$ or $\hat{A}_{m0}\hat{A}_{0n}\hat{A}_{00}\hat{A}_{00}$ in Eq.(76).

Our next task is therefore to find the $1/\Omega$ scaling of the spectral weights $\hat{A}_{mn}$ and $\hat{A}_{m0}$. We do this on a model by model basis. We will only estimate the integrals in the power-counting sense.

**SYK$_{222}$:** We first consider the SYK$_{222}$ model with $t_L = t_R = t$, which is a quadratic theory. The Green's function and self-energies that appear below refer to the retarded component. To perform the $1/\Omega$ expansion, we write $\omega + \hat{m}\Omega - \hat{\Sigma}$ into a block matrix

$$\begin{pmatrix} A & B \\ C & D \end{pmatrix}, \tag{81}$$

where $A_{00} = \omega - \hat{\Sigma}_{00}$, $B_{0m} = -\hat{\Sigma}_{0m}$, $C_{m0} = -\hat{\Sigma}_{m0}$, $D_{mn} = (D_0)_{mn} - \hat{\Sigma}_{mn}$, $(D_0)_{mn} = (\omega + m\Omega)\delta_{mn}$. Using block matrix inversion identities, we can write

$$\hat{G} = \begin{pmatrix} \frac{1}{A - BD^{-1}C} & -\frac{1}{A - BD^{-1}C} B \frac{1}{D} \\ -\frac{1}{D - CA^{-1}B} C \frac{1}{A} & \frac{1}{D - CA^{-1}B} \cdot \end{pmatrix} \tag{82}$$

We can now solve the Schwinger dyson equation order by order in $1/\Omega$.

We will first look for nontrivial corrections arising from $\hat{G}_{mn}$ with $m, n \neq 0$, and then those from $\hat{G}_{mn}$ with $m \neq 0$. At the $1/\Omega^0$ order, only the 00th block of the Green's function $\hat{G}_{00}$ and the self-energy $\hat{\Sigma}_{00}$ are nonzero.

Going to first order in $1/\Omega$, there is a non-zero term in $\hat{G}_{mn}$:

$$\hat{G}_{\alpha,mn}^{(1)} = \frac{1}{\omega + m\Omega}\delta_{mn}\,. \tag{83}$$

However, this term does not contribute to the spectral weight because it has no discontinuity across the real axis for $-\Omega/2 < \omega \leq \Omega/2$.

Therefore we need to go to the next $1/\Omega^2$ order, where the Green's function reads

$$\hat{G}_{\alpha,mn}^{(2)} = \frac{1}{\omega + m\Omega}\left[\hat{\Sigma}_{\alpha,mn} + \hat{\Sigma}_{\alpha,m0}\hat{G}_{\alpha,00}\hat{\Sigma}_{\alpha,0n}\right]^{(0)} \frac{1}{\omega + n\Omega}\,. \tag{84}$$

This needs to be solved together with the self-energy, which reads

$$\hat{\Sigma}_{\alpha,mn} = t^2 \hat{G}_{\alpha,mn} + V_{LR}^2 \sum_{ab} F_{m-a} F_{n-b}^* \hat{G}_{\bar{\alpha},ab}. \tag{85}$$

For the result to be second order, the only possibility is to pick up the zeroth order $a = b = 0$ term in the sum, therefore we have

$$\hat{G}_{\alpha,mn}^{(2)} = \frac{V_{LR}^2 F_m F_n^* \hat{G}_{\bar{\alpha},00}}{(\omega + m\Omega)(\omega + n\Omega)}\,, \quad \hat{A}_{\alpha,mn}^{(2)} = \frac{V_{LR}^2 F_m F_n^* \hat{A}_{\bar{\alpha},00}}{(\omega + m\Omega)(\omega + n\Omega)}\,. \tag{86}$$

However, this term does not contribute to the conductivity at the order $1/\Omega^2$, this is due to cancellation between $m > 0 (n > 0)$ and $m < 0 (n < 0)$ terms in the sum (74):

$$\text{Re}\,\sigma(\omega) \propto \sum_{mn} F_{-m} F_{-n}^* \hat{P}_{LR,mn}(\omega) \sim \sum_{mn} F_m F_n^* \hat{A}_{L,mn}^{(2)} \hat{A}_{L,00}$$
$$\sim \sum_{m,n} \frac{F_m F_{-m}}{\omega + m\Omega} \frac{F_n^* F_{-n}^*}{\omega + n\Omega} \hat{A}_{L,00}^2 + (L \leftrightarrow R)\,. \tag{87}$$

Here for clarity we have dropped the other factors and integration on frequencies. As we see $\sum_m 1/(\omega + m\Omega) \sim \omega/\Omega^2$ due to cancellation between $m$ and $-m$, Therefore, the contribution to conductivity is actually of order $1/\Omega^4$.

Moving down to order $1/\Omega^3$, the Green's function becomes

$$\hat{G}_{mn}^{(3)} = \frac{1}{\omega + m\Omega} \left[ \hat{\Sigma}_{\alpha,mn} + \hat{\Sigma}_{\alpha,m0} \hat{G}_{\alpha,00} \hat{\Sigma}_{\alpha,0n} \right]^{(1)} \frac{1}{\omega + n\Omega}$$
$$+ \sum_{k \neq 0} \frac{1}{\omega + m\Omega} \left[ \hat{\Sigma}_{\alpha,mk} + \hat{\Sigma}_{\alpha,m0} \hat{G}_{\alpha,00} \hat{\Sigma}_{\alpha,0k} \right]^{(0)} \frac{1}{\omega + k\Omega} \left[ \hat{\Sigma}_{\alpha,kn} + \hat{\Sigma}_{\alpha,m0} \hat{G}_{\alpha,00} \hat{\Sigma}_{\alpha,0n} \right]^{(0)} \frac{1}{\omega + n\Omega} \,, \tag{88}$$

where ($m$) in the superscript means taking the $1/\Omega^m$ order term of the bracket. There is also no valid contribution as this order. Looking at the first line of (88), the first order term in the bracket is from $\hat{\Sigma}_{\alpha,mn}^{(1)}$:

$$\hat{\Sigma}_{\alpha,mn}^{(1)} = \frac{t^2}{\omega + m\Omega} \delta_{mn} + V_{LR}^2 \sum_{a \neq 0} F_{m-a} F_{n-a}^* \frac{1}{\omega + a\Omega} \,, \tag{89}$$

which is non-singular and does not contribute to spectral weight. The second term $\hat{\Sigma}_{m0} \hat{G}_{00} \hat{\Sigma}_{0n}$ does not have first order term due to $F_0 = 0$. As for the second line of (88), it suffers the same cancellation as described above in the $k$-summation and also later in the $m,n$-summation, rendering this contribution at least of order $1/\Omega^6$. Therefore, the first non-trivial correction to $\hat{G}_{\alpha,mn}$ appears at the order of $1/\Omega^4$.

Next, we investigate the first non-trivial correction in $\hat{G}_{0m}$ with $m \neq 0$. According to Eq.(82), it has the form

$$\hat{G}_{\alpha,0m} = \hat{G}_{\alpha,00} \sum_{n \neq 0} \hat{\Sigma}_{\alpha,0n} \hat{G}_{\alpha,nm} \,. \tag{90}$$

We now argue that $\hat{G}_{\alpha,0m}$ starts at least from order $1/\Omega^2$. Since $\hat{G}_{\alpha,nm}$ is at least first order, we just need to analyze $\hat{\Sigma}_{\alpha,0n}$ and show its first order. Indeed,

$$\hat{\Sigma}_{\alpha,0n} = t^2 \hat{G}_{\alpha,0n} + V_{LR}^2 \sum_{ab} F_{m-a} F_{-b}^* \hat{G}_{\bar{\alpha},ab} \,, \tag{91}$$

and both terms are at least of order $1/\Omega$ (note that $b \neq 0$ due to freezing). Therefore $\hat{A}_{\alpha,0m}$ is at least of order $1/\Omega^2$. Since it contributes to $\hat{P}_{mn}$ in the form of $\hat{A}_{m0} \hat{A}_{0n}$, it contributes to the conductivity at the order of $1/\Omega^4$.

Combining everything, we conclude that the inter-dot conductance of SYK$_{222}$ is of the form

$$\text{Re}\,\sigma(0) = \frac{V_{LR}^2 t^2}{\Omega^4} \mathcal{F}_{222}(V_{LR}/t, \beta_{\text{eff}} t) \,, \tag{92}$$

where $\mathcal{F}_{222}$ is a dimensionless scaling function.

**SYK$_{242}$:** In this part, we perform a similar analysis for the SYK$_{242}$ model. The second order contribution to $\hat{G}_{mn}$ is still given by Eq.(84). The difference is now the self-energy is instead given by

$$\hat{\Sigma}_{\alpha,mn} = t^2 \hat{G}_{\alpha,mn} + J_{LR}^2 \sum_{ab} F_{m-a} F_{n-b}^* \hat{M}_{ab} \,, \tag{93}$$

where $\hat{M}$ is the product of three Green's functions:

$$M_{\alpha}^{\gtrless}(t,t') = G_{\bar{\alpha}}^{\gtrless}(t,t')^2 G_{\alpha}^{\lessgtr}(t',t) \,. \tag{94}$$

Its spectral function is given by (after applying the generalized Kadanoff-Baym ansatz)

$$\hat{M}_{\alpha,mn}^{>}(\omega) - \hat{M}_{\alpha,mn}^{<}(\omega) = i \sum_{m_1,m_2,n_1,n_2} \int_{-\Omega/2}^{\Omega/2} \frac{d\omega_1 d\omega_2}{(2\pi)^2} \hat{A}_{\alpha,m_1 n_1}(\omega_1) \hat{A}_{\bar{\alpha},m_2 n_2}(\omega_2) \hat{A}_{\alpha,m_1 + m_2 - m, n_1 + n_2 - n}(\omega_1 + \omega_2 - \omega)$$
$$\times \frac{-\cosh(\beta_{\text{eff}} \omega/2)}{4 \cosh(\beta_{\text{eff}} \omega_1/2) \cosh(\beta_{\text{eff}} \omega_2/2) \cosh(\beta_{\text{eff}}(\omega_1 + \omega_2 - \omega)/2)} \,. \tag{95}$$

Following the same previous reasoning, zeroth order term of $\hat{\Sigma}_{\alpha,mn}$ is

$$\hat{\Sigma}_{\alpha,mn}^{(0)} = J_{LR}^2 F_m F_n^* \hat{M}_{00} \,. \tag{96}$$

However this contribution suffers a similar cancellation to Eq.(87):

$$
\begin{aligned}
\mathrm{Re}\,\sigma(\omega) &\propto \sum_{mn} F_{-m} F_{-n}^{*} \hat{Q}_{LR,mn}(\omega) \sim \sum_{mn} F_{-m} F_{-n}^{*} \hat{A}^{(2)}_{L,mn} \hat{A}_{L,00} \hat{A}_{R,00} \hat{A}_{R,00} \\
&\sim \sum_{m,n} \sum_{\bar m,\bar n} \frac{F_m F_{-m}}{\omega + m\Omega} \frac{F_{\bar n}^{*} F_{-n}^{*}}{\omega + n\Omega} i(\hat{M}^{>}_{L,00} - \hat{M}^{<}_{L,00}) \hat{A}_{L,00} \hat{A}_{R,00} \hat{A}_{R,00} + (L \leftrightarrow R)\,,
\end{aligned}
\tag{97}
$$

which makes its contribution to conductance of order $1/\Omega^4$.

We move to the next order $1/\Omega^3$. Unlike the SYK$_{222}$ case, the there is actually a non-trivial contribution to $\hat{G}_{mn}$ appearing at this order for SYK$_{242}$. In SYK$_{222}$, since $\hat{M}$ is linear in $\hat{G}$, the first order term $\hat{M}^{(1)} \propto G^{(1)}_{mn}$ is non-singular and does not contribute to spectral weight (see the second term of Eq.(89)). In contrast, in SYK$_{242}$, due to the nonlinearity of $\hat{M}$, there is a non-trivial first order spectral weight. The explicit expression is

$$
\begin{aligned}
\hat{M}^{>,(1)}_{\alpha,mn}(\omega) - \hat{M}^{<,(1)}_{\alpha,mn}(\omega) = \int_{-\Omega/2}^{\Omega/2} \frac{d\omega_1 d\omega_2}{(2\pi)^2} &\Big[ \hat{A}_{\bar\alpha,00}(\omega_1) \hat{A}_{\bar\alpha,00}(\omega_2) \hat{G}^{(1)}_{\alpha,-m,-n}(\omega_1 + \omega_2 - \omega)(n_F(\omega_1) + n_F(\omega_2) - 1) \\
&+ 2\hat{G}^{(1)}_{\bar\alpha,mn}(\omega_1) \hat{A}_{\bar\alpha,00}(\omega_2) \hat{A}_{\alpha,00}(\omega_1 + \omega_2 - \omega) n_F(\omega_1 + \omega_2 - \omega) \Big]\,,
\end{aligned}
\tag{98}
$$

where

$$
\hat{G}^{(1)}_{\alpha,mn} = \frac{1}{\omega + m\Omega} \delta_{mn}\,.
\tag{99}
$$

This term can lead to the following spectral weight of $\hat{A}_{mn}$ via the first line of Eq.(88):

$$
\hat{A}^{(3)}_{\alpha,mn} = \frac{V_{LR}^2}{(\omega + m\Omega)(\omega + n\Omega)} \sum_{ab} F_{m-a} F_{n-b}^{*} i\left(\hat{M}^{>,(1)}_{ab} - \hat{M}^{<,(1)}_{ab}\right)\,.
\tag{100}
$$

The discussion for the other type of correction, $\hat{A}_{m0}$, is similar to SYK$_{222}$. We find that $\hat{A}_{m0}$ starts from order $1/\Omega^2$.

Therefore, in a non-quadratic system like SYK$_{242}$, we find that the inter-dot conductance is proportional to $1/\Omega^3$, and it can be written in the form

$$
\mathrm{Re}\,\sigma(0) = \frac{J_{LR}^2 t}{\Omega^3} \mathcal{F}_{242}(J_{LR}/t, \beta_{\mathrm{eff}} t)\,,
\tag{101}
$$

with a dimensionless scaling function $\mathcal{F}_{242}$.

**SYK$_{442}$:** This case is similar to SYK$_{242}$. The difference now is that there can be an additional $1/\Omega^3$ contribution coming from the intra-dot interaction of the $L$ dot. The inter-dot conductance is of the form

$$
\mathrm{Re}\,\sigma(0) = \frac{J_{LR}^2 t_R}{\Omega^3} \mathcal{F}_{442}(J_{LR}/t_R, J_L/t_R, \beta_{\mathrm{eff}} t_R)\,.
\tag{102}
$$

### Decay rate of density difference and comparison with numerics

Finally, let us convert the conductivity calculated above to a decay rate of density difference, and compare it with the numerical results. By definition, the conductivity is the response of current to the difference in electrostatic potential

$$
\dot n_L = -\dot n_R = \sigma(V_R - V_L)\,.
\tag{103}
$$

To get a closed equation for $n_\alpha$ and $V_\alpha$, we need to know the compressibility of the two dots

$$
\kappa_\alpha = \frac{\partial n_\alpha}{\partial V_\alpha}\,.
\tag{104}
$$

At freezing, the two dots are approximately decoupled, and we expect the compressibility takes the value of the individual isolated dots. Therefore, the time-evolution for the potential difference is

$$
\frac{d(V_R - V_L)}{dt} = -\lambda(V_R - V_L)\,,
\tag{105}
$$



\end{document}